# Title:
# Switch-like enhancement of epithelial-mesenchymal transition by YAP through feedback regulation of WT1 and small Rho-family GTPases

**Authors:**

JinSeok Park[1#], Deok-Ho Kim[2#], Sagar R. Shah[3,4], Hong-Nam Kim[5], Kshitiz[1], David Ellison[3], Peter Kim[2], Kahp-Yang Suh[4], *Alfredo Quiñones-Hinojosa*[4*], *Andre Levchenko*[1*]

[#]These authors contributed equally to this work

**Affiliation:**

[1]Department of Biomedical Engineering and Yale Systems Biology Institute, Yale University, New Haven, CT 06520, USA.

[2]Department of Bioengineering, University of Washington, Seattle, WA 98195, USA.

[3]Department of Biomedical Engineering, The Johns Hopkins University School of Medicine, Baltimore, MD 21231, USA.

[4]Department of Neurologic Surgery, Mayo Clinic, Jacksonville, FL 32224, USA.

[5]Department of Mechanical & Aerospace Engineering, Seoul National University, Seoul, Republic of Korea

*To whom correspondence should be addressed: Alfredo Quiñones-Hinojosa, M.D. (Quinones-Hinojosa.Alfredo@mayo.edu), Andre Levchenko, Ph.D. (andre.levchenko@yale.edu)

## ABSTRACT

Collective cell migration is a hallmark of developmental and patho-physiological states, including wound healing and invasive cancer growth. The integrity of the expanding epithelial sheets can be influenced by extracellular cues, including cell-cell and cell-matrix interactions. We show the nano-scale topography of the extracellular matrix underlying epithelial cell layers can have a strong effect on the speed and morphology of the fronts of the expanding sheet triggering epithelial-mesenchymal transition (EMT). We further demonstrate that this behavior depends on the mechano-sensitivity of the transcription regulator YAP and two new feedback cross-regulation mechanisms: through Wilms Tumor-1 and E-cadherin, loosening cell-cell contacts, and through Rho GTPase family proteins, enhancing cell migration. These

YAP-dependent regulatory feedback loops result in a switch-like change in the signaling and expression of EMT-related markers, leading to a robust enhancement in invasive epithelial sheet expansion, which might lead to a poorer clinical outcome in renal and other cancers.

**INTRODUCTION**

Reorganization of epithelial layers is a major hallmark of developmental processes, as well as physiologically important events such as wound healing and tumorigenesis [1-3]. Destabilization of epithelial layers and violation of their integrity can jeopardize the function of the tissue. Nevertheless, epithelial organization has to undergo transient disintegration during both normal and pathological instances of the epithelial-to-mesenchymal transition (EMT), e.g., in the formation of migrating neural crest cells or initiation of cancer metastasis [4-6].

Although the analysis of migrating epithelial sheets both *in vivo* and in the model tissue culture environments has yielded many insights into the mechanisms underlying collective cell behavior, much remains poorly understood. For instance, recent results indicate that the organization and composition of the extracellular matrix (ECM) can dramatically change the collective behavior of mammary epithelial cells and increase the propensity for invasive and metastatic behavior of mammary tumors [7-9]. This finding, along with accumulating evidence of the crucial role of ECM re-organization in invasive cancer growth [10-12], strongly suggests that ECM organization can be an important regulator of both individual and collective cell migration. Furthermore, ECM re-organization also controls EMT-like processes in various developmental settings, e.g., in gastrulation and the onset of neural crest cell migration [13,14], as well as in would healing [15]. Therefore, it is important to explore how the organization of the ECM underlying epithelial sheets could affect their migratory responses, and whether this organization can challenge the integrity of the epithelial layers and promote EMT.

To address the effects arising from frequently anisotropic and aligned organization of ECM, here we used topographically structured cell adhesion substrata with nano-scale aligned fibers similar in size to those found in ECM. Such nano-textured cell adhesion substrata have been extensively used to mimic individual cell migration on aligned ECM fibers, since they are structured to resemble these fibers in size, chemical composition and spatial orientation [16-18]. The movement of cells on such surfaces has displayed many important similarities to migration patterns frequently observed *in vivo* [19,20].

In this study, we observed that topographic organization of the cell adhesion substratum mimicking ECM

fiber alignment triggered EMT-like processes in the expanding epithelial cell sheets. Furthermore, we provide evidence that these properties of normal epithelial cells are in large part due to an all-or-none, switch-like activation of YAP (Yes-associated protein), a transcriptional co-regulator previously implicated in the control of cell-cell and cell-substratum interactions, and in mechanosensing [21-25]. We delineate the molecular mechanisms involved in this switch-like response, demonstrating that it depends on feedback interactions, involving Wilms Tumor-1 (WT1)-dependent YAP crosstalk with E-cadherin, as well as the Merlin-dependent YAP crosstalk with Rho-family small GTPases, uncovering previously unknown regulatory effects of YAP. Overall, our results strongly argue that YAP-mediated, switch-like responses to changes in ECM organization can be the key factor determining the integrity of epithelial sheets and the onset of EMT, both in health and disease, and suggest the mechanisms controlling this collective cell response.

RESULTS

**Mechanical cues mimicking aligned ECM topography enhance the EMT-like phenotype**

We set out to explore the putative mechanistic underpinnings of the effects of the nano-topographic ECM structure by culturing cells on arrays of nano-scale aligned ECM fiber-like ridges coated with Collagen type I. This *in vitro* model has been shown to mimic multiple aspects of the natural ECM structure and control various features of single and collective cell behavior[16]. Consistent with prior observations[26], epithelial sheets of Madin-Darby Canine Kidney (MDCK) cells displayed a 50% increase in the velocity and persistence of cell migration in the direction of these Nano-structured Ridge Arrays (NRA) vs. the control flat surfaces (Figs. 1a,b, Supplemental figure 1 and Supplemental video 1). The cell migration speed in NRA was maximal at the edges of expanding sheets, gradually decreasing for cells deeper in the sheet, in contrast to the more even speed distribution on the flat surfaces (Fig. 1c and Supplemental figure. 1). The edges of the sheets also underwent a dramatic increase in the formation of finger like projections (FLPs), with the cells at the tips of the projections displaying distinct cytoskeleton structure (Supplemental figures 2,3 and Supplemental videos 2,3) and frequent separations from the rest of the FLPs, the behavior resembling partial and complete EMT (Fig. 1d, Supplemental figure 4a, and Supplemental video 4).

On the molecular level, EMT is frequently characterized by expression and localization of well-established markers, such as β-catenin, Snail, Slug, Twist and vimentin[4]. Indeed, we found higher expression of Snail in cells cultured on NRA vs. flat surfaces (Fig. 1e) and increased expression of Slug and Twist in FLPs at the edge of cell sheet on NRA surfaces, but not at the sheet edges on flat surfaces (Supplemental figure 4b). Vimentin also displayed a strongly enhanced expression on NRAs, with the expression being particularly elevated in FLPs (Supplemental figure 4c). These results provided further support for the EMT-like nature of the epithelial cell expansion on NRA, particularly at the FLP structures at the edges of the expanding sheets.

To test whether the effects of the substratum nano-topography would be synergistic with the effects of more established EMT inducers, such as TGFβ[27], we examined the outcomes of TGFβ stimulation on the surfaces of distinct topographies. We found that, although, as expected, the number of cell dissemination events was greatly increased in the presence of TGFβ on the flat surfaces, the effect was similar to the effect of substratum topography cue alone. Strikingly, when TGFβ and nanotopographic cues were combined, the number of dissemination events increased 4–fold vs. the effect of each of these cues alone (Fig. 1f), indicating a strong synergy between these two inputs. Overall, these data argued that an EMT-like process could be induced by nanotopographic cues in a fashion distinct from but synergistic with TGFβ, leading increase in expression of EMT markers

**YAP is a critical switch-like mediator of the enhanced EMT-like phenotype in response to mechanical cues**

What is the mechanism of the dramatic increase in the probability of the EMT-like events at the edge of the epithelial monolayers expanding on NRA? We focused on the analysis of YAP, since this transcriptional regulator has been implicated both as a mechanosensor[22] and as a regulator of the balance between cell-cell[28] and cell-ECM interactions[21]. It is also known to promote wound healing[29] and certain EMT-like features[30]. Indeed, we found that cells on NRA surfaces exhibited decreased phosphorylation of YAP indicating high propensity of nuclear translocation of YAP compared to cells cultured on flat substrata (Fig. 1g). Disruption of YAP expression by RNA interference substantially reduced the speed of cell migration (Fig. 1h and Supplemental video 5) and the extent of FLP formation, essentially to the levels seen in the unperturbed cells on flat substrata, suggesting that YAP activity indeed mediated the mechanosensing-dependent aggressive epithelial monolayer expansion.

When examined on the single cell level, the localization of YAP was strikingly distinct in cells cultured on different substrata. Whereas it was exclusively cytosolic in the cells cultured on flat substrata, it displayed an essentially all-or-none nuclear/cytosolic distribution on NRA surfaces — changing from mostly nuclear in the FLPs to mostly cytosolic in the bulk of the monolayer (Fig. 1i). This change in YAP localization was sharp, with no cells displaying YAP localization to both cytosolic and membrane domains, as captured by a bi-modal distribution (Fig. 1j). In addition, the intracellular localization patterns of active β-catenin, another EMT marker, incorporating with YAP during its regulation in advancing MDCK sheets were very similar with YAP localization: entirely plasma membrane-associated active β-catenin in cells on the flat substrata and its variation from mostly nuclear in the finger-like structures at the edge of the expanding layers to mostly plasma membrane-localized in the cells in the bulk of the layer with switch-like bi-modal fashion (Figs. 1i,j). Furthermore, knockdown of YAP expression led to a reversal of localization of active β-catenin to the plasma membrane in virtually all cells (Supplemental figure 5a). Staining for vimentin also suggested substantial reduction of its expression in YAP$^{KD}$ cells vs. the levels seen in unperturbed cells (Supplemental figure 5b), when both were cultured on topographically defined substrata. These results strongly suggested the critical nature of YAP signaling in inducing EMT on the aligned fibrous cell adhesion substrata. We next explored the mechanisms of the switch-like YAP activation and its effect on this EMT model.

**YAP can control mechanically induced EMT through feedback regulation of E-cadherin via WT1 influencing cell-cell coupling**

Loosening of cell-cell contacts is thought to be a necessary step in ensuring effective EMT. Therefore, we examined whether mechanically induced YAP activity could affect the critical mediator of epithelial cell contacts, E-cadherin. We found that inhibition of E-cadherin-mediated cell-cell interaction led to a profound increase in cell dissemination, with the speed of individual cells and small cell clusters approximately equal to speed of the single cells undergoing dissemination under unperturbed (control) conditions on NRA (Fig. 2a and Supplemental video 6). Importantly, silencing of YAP expression led to a much weaker effect of E-cadherin inhibition, suggesting that YAP can negatively control E-cadherin mediated cell-cell contact. Indeed, we observed not only a substantial increase in E-cadherin expression in YAP$^{KD}$ cells, accompanied by decreased activation of β-catenin, but also decrease in E-cadherin expression and increased activation of β-catenin in cells with increased levels of YAP expression (YAP$^{OE}$) cells (Fig. 2b).

To further explore the mechanistic details of the putative YAP-E-cadherin interaction, we examined the known suppressor of E-cadherin expression, the Wilms Tumor protein (WT1)$^{31,32}$. This protein is particularly interesting to evaluate due to its role in regulating mesenchymal-epithelial transition (MET) and cell-cell interactions in developing kidney (making MDCK cells a relevant cell type model) and associated malignancies. Surprisingly, we found that that WT1 localization was very similar to the nuclear and cytoplasmic YAP localization patterns in various locations across the expanding epithelial layer (Fig. 2c). Furthermore, silencing of YAP expression led to a decrease in nuclear localization of WT1 (Fig. 2d). Moreover, we found that WT1 and YAP displayed similar variation of nuclear localization patterns as a function of cell density (Figs. 2e,f). These results suggest that these two proteins can structurally and functionally interact in MDCK cells, the hypothesis that we then explored in detail.

We observed that, $YAP^{KD}$ cells showed no nuclear localization of WT1 on NRA, suggesting that YAP can facilitate transport of WT1 into the nucleus (Fig. 3a). Conversely, $WT1^{KD}$ cells under the same conditions still displayed nuclear YAP, suggesting that YAP localization patterns are not affected by WT1, and thus YAP can be the vehicle enhancing nuclear WT1 localization, but not vice versa. Furthermore, co-immunoprecipitation experiments indicated WT1-YAP complex formation raising the possibility that these two transcriptional regulators can jointly control gene expression, potentially linking YAP to E-cadherin control through WT1 co-regulation (Fig. 3b). To explore whether WT1-YAP complex directly controls E-cadherin expression, we performed chromatin immunoprecipitation (ChIP) with YAP and WT1 antibodies at E-cadherin promoter sequence[33]. This finding confirmed that WT1-YAP complex binds directly to the E-cadherin promoter sequence and can thus control E-cadherin gene expression (Fig. 3c). Furthermore, we found that WT1 knockdown did lead to an increase in E-cadherin expression in MDCK cells, supporting a negative effect of WT1 on E-cadherin expression consistent with the effect of YAP knockdown (Fig. 3d). Although the knockdown of WT1 itself did not change cell migration speed in cell sheets cultured on NRA, it desensitized cells to E-cadherin inhibition, restoring their migration and dissemination responses to those of the control cells (Figs. 3e,f and Supplemental video 7), further suggesting that the effect of WT1 was through regulation of E-cadherin-mediated cell junctions. These results were similar to the effects of YAP silencing, supporting functional coupling between YAP and WT1 and implying that E-cadherin effects on cell migration were YAP and WT1 dependent. As E-cadherin blocking antibody also activated YAP (Fig. 3g), our results suggested a double negative feedback, involving E-cadherin and YAP-WT1 complex formation, leading to switch-like enhanced loosening of cell contacts and increased cell dissemination from FLPs in cell sheets cultured on NRAs (Fig. 3h).

**YAP can control mechanically induced EMT through regulation of small Rho-family GTPases via Merlin influencing cell speed**

In addition to loosening cell-cell adhesions, WT1-independnet reduction of the speed of cell migration following YAP knockdown suggested that YAP activity might directly impinge on the regulation of cell polarity and migration. We found that cells cultured on NRA displayed enhanced activity of a small Rho-family GTPase, Rac1, implicating in regulation of cell migration[34-36] (Fig. 4a). Furthermore, silencing YAP expression abolished Rac1 activation and conversely, increasing YAP expression led to increased Rac1 activation, suggesting a key role of YAP in regulation of Rac1 activity on NRA (Fig. 4b). YAP knockdown also downregulated the expression of TRIO, a Rac1 activating GTP exchange factor and the Rac1 effector p21-actiated kinase (PAK), whereas YAP overexpression had the opposite effect, further suggesting that YAP-mediated Rac1 regulation is functionally significant (Fig. 4b). Inhibition of TRIO phenocopied inhibition of Rac1 activity in cell migration assay (led to suppression of enhanced motility) suggesting that YAP can control Rac1 activation and cell migration, at least in part, through regulation of TRIO expression (Fig. 4c and Supplemental video 8 and 9, also see the accompanying paper, by Shah *et al.*, demonstrating TRIO regulation by YAP in other cell types).

We found that YAP silencing down-regulated Merlin showing its decreased phosphorylation (Fig. 4b), a negative regulator of YAP[37] also known to control and be controlled by Rac1 and PAK activity, consistent with prior reports indicating that enhanced YAP up-regulates Merlin[38,39]. Potential feedback cross-regulation between YAP and Merlin suggested a mechanism of YAP-Rac1 feedback interaction involving PAK that can be enhanced in cells incubated on NRAs. To test this possibility further, we confirmed that culturing cells on NRA (but not flat surfaces) decreased association of Merlin with cell-cell junctions, particularly in FLPs (Fig. 4d), consistent with reduced association of Merlin with angiomotin (AMOT), a plasma membrane bound protein localized to cell-cell junctions (Fig. 4e). This result was supported by down-regulation of Merlin phosphorylation in $YAP^{KD}$ cells and up-regulation of Merlin phosphorylation in $YAP^{OE}$ cells (Fig. 4b), as well as increased association of Merlin with AMOT in $YAP^{KD}$ cells (Fig. 4f), in agreement with the prior results suggesting that a decrease in PAK-mediated phosphorylation functionally suppresses Merlin through controlling its association with AMOT[40-42].

Importantly, silencing of Merlin expression both increased Rac1 activity and decreased phosphorylation (and thus increased the activity) of YAP (Fig. 4g). Consistent with these effects, decreased Merlin expression led to a significant increase in cell migration speed, which was reversed by Rac1 inhibition (Fig. 4h and Supplemental video 10). Merlin silencing also led to a significantly greater occurrence of cell dissemination events (Fig. 4i). These results suggested that Merlin contributes to Rac1 regulation in a feedback fashion, by negatively controlling YAP, which in turn negatively controls Merlin, in a Rac1 and PAK dependent fashion, in cells cultured on NRA (Fig. 4j).

Antagonistic interaction between Rac1 and RhoA small GTPases is thought to be critical for cell polarization and optimal cell migration[36,43]. We found an increase in MLC phosphorylation following silencing of YAP expression, which was particularly enhanced at the sheet edges (Supplemental figure 6a), suggested that YAP can regulate MLC kinase (MLCK) through regulation of the Rho-family small GTPases[44,45]. Indeed, we found that a knockdown of YAP led to enhanced MLC phosphorylation (Supplemental figure 6b), but, in contrast to a recent report[21], did not affect its overall expression level. Furthermore, we found that suppression of cell migration due to YAP knockdown was rescued by inhibition of the activity of Rho-dependent kinase (ROCK) in cells cultured on NRA (Supplemental figure 6c and Supplemental video 11). This result suggested that YAP activation in the edge cells cultured in the presence of topographic cues controls cell migration velocity through suppression of RhoA dependent signaling, by up-regulating Rac1 and down-regulating RhoA, either directly or through mutual inhibition of these small GTPases. Interestingly, in the absence of YAP silencing, the effect of ROCK inhibition on the cell speed in cells cultured on NRA was limited, suggesting that YAP activity can suppress ROCK very effectively in response to the mechanical cue.

**Feedback interactions involving YAP can promote switch-like onset of EMT**

The switch-like, bimodal nuclear translocation of YAP, WT1 and β-catenin (Figs. 1i,j and 2c) in the marginal and sub-marginal zones of expanding sheets is suggestive of underlying bi-stability in regulation of these proteins, which in turn is frequently associated with positive or double negative feedback interactions[44,46,47]. Indeed, our data suggest YAP can negatively control E-cadherin expression through interaction with WT1 (Fig. 3h). This finding coupled with prior observations that E-cadherin can negatively control YAP activity[28], and our finding that inhibition of E-cadherin leads to a progressive decrease in phosphorylation of YAP, i.e., increase in YAP activity (Fig. 3g), provides evidence for a double negative feedback between YAP and E-cadherin. In addition, our results reveal that YAP can be a part of a positive feedback involving Merlin and the Rho family small GTPases, particularly Rac1 (Fig. 4j). This combination of positive (YAP and Rac1) and double negative (YAP and E-cadherin) feedback interactions triggered or enhanced by the mechanical cue can lead to a switch-like enhancement of the EMT phenotype, which indeed was supported by a simple mathematical model, accounting for the underlying bi-stability of YAP-dependent signaling networks (Fig. 5a and Supplemental discussion). This model provided excellent fit to the bi-modal distributions of nuclear localization of YAP at different distances from the sheet edge, placing maximum bi-modality at the bases of forming FLPs, and providing a mechanism for instability at the sheet fronts leading to FLPs formation. To validate this model, we sought to vary the mechanical cue stimulating the YAP activity. We achieved this by further modifying NRA substrata by using materials having differing stiffness. YAP is known to be responsive to this type of mechanical cue, which was evidenced by stiffer NRA substrata leading cells to establish more aligned FAs (Supplemental figure 7) and steeper gradients of the cell speed values (Fig. 5b). The mathematical model suggested that decreasing substrate rigidity can lead to a gradual shift of YAP activation to the edge cells, thus decreasing the number of cells with the fully active YAP (Fig. 5c). These results were indeed supported by the experimental analysis (Fig. 5d), suggesting that both topographic structure and effective stiffness of the underlying ECM can control YAP-dependent epithelial edge expansion and propensity to undergo EMT (Supplemental figure 8 and Supplemental discussion).

**Clinical relevance of mechanically induced EMT-like behavior to renal cancer**

To examine the relevance of our findings to renal cancers that may best correspond in their origin to our model kidney cell line, MDCK, we first explored the behavior of a renal cancer cell line, ACHN. When cultured on NRA, the cells displayed the behavior similar to that of MDCK, including EMT-like phenotype and cell dissemination. We then contrasted the expression levels of various molecules implicated in the EMT of MDCK cells in disseminated ACHN cells vs. the cells remaining in the monolayer. We found the results to be fully consistent with those displayed by the MDCK cells, for 8 diverse proteins indicative of EMT and its feedback regulation by WT1 and TRIO (Fig. 6a). We then analyzed the patient data, querying the sequencing and expression profiles for 451 renal cancer samples contained in the TCGA database. We found that YAP1 and WT1 alterations were present in 5-6% of the cases, showing mostly non-overlapping patterns of alterations (Fig. 6b). Since our analysis suggests that these factors may control the expression of E-cadherin and other proteins as co-factors, we expected that altered expression of either one of them can be predictive of poor clinical outcome, with the joint YAP1-WT1 signature increasing the predictive power of the analysis. We indeed found that alterations in YAP1 led to a much worse prognosis, with the predictive power gaining much higher significance when both YAP1 and WT1 alterations were included in the signature (Fig. 6c).

**DISCUSSION**

Migration of extending epithelial sheets strongly depends on various chemical and mechanical environmental cues present in extracellular milieu[1,2,48]. Although re-organization of extracellular matrix, particularly an increase in the anisotropy of the fiber orientation has been linked to invasive cell behavior, the role of this process in EMT has not been well established. Our findings provide evidence for the important role of anisotropic nano-scale texture of the substratum approximating the structure of large ECM fibers in regulating EMT, and enhancing the collective invasive properties of cells at the edges of the expanding epithelial sheets. This mechanical cue has the strong and switch-like effect on EMT, similar in magnitude to the more established chemical inputs, such as TGFβ, and allows for strong synergy with this and, potentially, other chemical factors. This result underscores the importance of environmental inputs in control of cell invasiveness, which can be strongly enhanced even in the absence of genetic perturbations.

The mechanism of EMT regulation by the topography of the cell adhesion substratum suggested by the results presented here assigns a central role to feedback interactions involving a transcriptional co-regulator, YAP. This molecule was previously implicated both as mechanosensor and a regulator of cell-cell vs. cell-matrix interactions[21,22,28]. We identified two feedback interactions through which YAP can control the EMT-like phenotype in response to nanotopographic, mechanical stimulation. These feedback interactions suggest simultaneous suppression of E-cadherin expression through a WT1-dependent transcriptional regulation leading to disruption of the integrity of the epithelial sheet, and enhancement of Rac1 activity and thus cell migration speed in Merlin and small GTPase-dependent fashion. These findings and the findings of the accompanying paper (Shah *et al.* included in Supplemental data) suggest that YAP activation can be a central regulator of both single and collective cell motility in the presence of complex mechanical cues supplied by the matrix, providing further insight into the general mechanisms of contact guidance. The combined effects of the two feedback loops create a robust switch that can be modulated by other environmental inputs, including the stiffness of the substratum, the cue that has been implicated in the invasive cancer spread. In particular, decreasing ECM rigidity might partially negate the EMT-inducing effect of aligned ECM fiber structure, providing a potential for new interventions into the aggressive spread of normal and cancerous cells.

Involvement of WT1-YAP interaction as a critical regulator of the mechanically induced EMT processes has a number interesting consequences for our understanding of developmental and pathogenic processes. First, since this protein has also been implicated as a crucial regulator of inverse process of mesenchymal-epithelial transition (MET), our results suggest that both EMT and MET processes can controlled by the same molecular network in a switch-like fashion, which might shed light on MET events during Wilms tumor progression and normal kidney development. In particular, they suggest a mechanistic explanation for the prior results implicating YAP in kidney development and for the consequences of YAP dysregulation in anaplastic Wilm's tumors[49,50]. Furthermore, our results can help elucidate the interplay between YAP and WT1, both implicated in the EMT-like events during epicardial development[51]. We also show that in contrast to the regulation of Merlin during collective cell migration on flat substrata[40], mutual regulation between YAP and Merlin becomes a critical part of enhanced and more individualized cell migration in the presence of additional mechanical inputs. The feedback nature of this interaction can help reconcile our results in this and companion paper (Shah *et al.*,) with the prior observations placing TRIO upstream of YAP[52]. In particular, our results suggest that TRIO can be thought of being both upstream and downstream of YAP, by virtue of being a part of the positive feedback linking YAP to small GTPases and cell migration. Overall, these mechanistic insights provide the framework for understanding the two components of a successful EMT process: increased cell migration and decreased cell linkage, through two mechanisms, both modulated by YAP and triggered by mechanical stimuli.

The feedback nature of EMT control can be critical both in normal processes involving EMT and in cancer progression. Our data suggest that the finding made in normal renal epithelial cells translate to renal cancer cells, permitting identification of a novel clinical signature, relying on the analysis of alterations in both YAP and WT1, which together account for a significant poor prognosis for patients in renal cancers. These results argue that the YAP-WT1 complex can be a critical predictor of renal cancer progression. We propose that the methods of analysis described in this study can be used to further investigate EMT in renal and other cancers and assist in furthering precision medicine approaches in the clinic.

Mechanical stimuli can be diverse, encompassing the topography, rigidity and other features emerging from particular ECM composition and organization. Our results suggest that YAP can integrate these diverse mechanical cues in controlling EMT, in a fashion explained by a simple mathematical model. Overall, our study suggests a sophisticated control of epithelial cell expansion and EMT onset in response to various cues, with potential relevance to diverse developmental, physiological and pathological processes. It elucidates how EMT can robustly occur in normal epithelial sheets following matrix re-organization, with a dramatic switch-like change in signaling profiles, enhancing cell speed and loosening cell-cell contacts, the processes that can be conserved in diverse physiological and pathological settings.

**EXPERIMENTAL PROCEDURES**

**Fabrication of the NRA substrata.** The nano-structured substrata (width: 800 nm, spacing: 800 nm, height: 800 nm) were fabricated by capillary force lithography with UV-curable polyurethane acrylate (PUA, Minuta Tech., South Korea) polymer having the rigidity value of 0.1 GPa, unless indicated otherwise. To fabricate a flexible polymer mold, a precursor of UV-curable polymer was drop-dispensed onto the pre-fabricated silicon master with large areas of $25 \times 25$ mm$^2$. Then a polyethylene terephthalate

(PET) (SKC Inc., South Korea) film with thickness of 50 μm was brought into conformal contact onto the polymer resin. After removing air bubbles with roller, UV ($\lambda$= 250 to 400 nm) was irradiated for few tens of seconds for the cross-linking. The flexible and transparent mold with PET support was obtained after peeling-off from the master, leaving behind regularly spaced nanogrooves, i.e., NRA. For the polymer nanogroove patterns on the slide glass-sized coverslips (50 x 24 mm, Fisher), the same replication process was performed onto a cleaned cover slip using the replicated PUA pattern as a mold. The flat surfaces were generated as a control experimental set with same procedure except no master. To study the effect of rigidity, we also used different pre-polymers for fabrication as follows: PU elastomer (MINS 311 RM) for 10 MPa was purchased from Minuta Tech. (Korea). Hard PU (NOA83 H) for 1 GPa was purchased from Norland Optical Adhesive Inc. (NY, USA). The detailed information of soft and intermediate PU materials can be found elsewhere[53]. These patters were assembled onto 4 multi-well chambers (Nunc® Lab-Tek® Chamber Slide™ system) after removing pre-mounted slide glasses on their bottom. We chose the dimensions of NRA based on the dimension that the most aggressive expansion was shown on variable-density ridged arrays. An example of an experiment examining differential speed regulation on such surfaces is shown in Supplemental figure 1.

**Analysis of epithelial sheet migration.** We purchased MDCK cells and GFP-α-tubulin and GFP-F-actin transfected MDCK cells from Marinpharm GmbH. Cells were cultured in Dulbecco's modified Eagle's medium (Gibco) supplemented with 10% fetal bovine serum (Gibco), 50 U ml$^{-1}$ penicillin, and 50 μg ml$^{-1}$ streptomycin (Invitrogen) at 37°C, 5% $CO_2$ and 90% humidity. These were split 1:4 after trypsinization for passaging every 2-3 days. Glass coverslips covered with the NRA nano-topographic substrata were pre-glued onto the bottom surface of the custom-made MatTek dish (P35G-20-C) or 4 multi-well chambers (Nunc® Lab-Tek® Chamber Slide™ system). Then, cells were re-plated on the patterns pre-coated by 30 mg/ml of collagen type I (Invitrogen, A1048301) for 3 hours and incubated for up to 12 hours for the generation of monolayer sheets. More specifically, on collagen type I pre-coated substrata, a PDMS insert (Supplemental figure 1a) that have two 7 x 2 mm rectangular holes was deposited and 2.0 x 10$^5$ cells in 50 μL of medium were seeded into each hole. After incubating for about 12 hours, the insert was gently removed with tweezers. For the perturbation of signaling pathway related to cell migration, 10 mM of Y27632 (Sigma-Aldrich), 100 mM of NSC 23766 (Tocris), 5 μg/mL of Mitomycin C (Sigma-Aldrich), 10 μM of ITX3 (Tocris), 10 μg/mL of E-cadherin antibody (Sigma-Aldrich, U3254) and 5 ng/mL of recombinant TGFβ1 (R&D Systems) were added. To show cell density-dependence of nuclear localization of YAP and WT1, 5 x 10$^5$ cells / cm$^2$ of cells were seeded for and the cell density were two-fold serially diluted three times. For immunofluorescence staining, the samples were chosen: dense one

($5 \times 10^5$ cells / cm$^2$) and sparse one ($1.25 \times 10^5$ cells / cm$^2$), and for immunobloting, we used all four different cell-density samples.

**Lentiviral transfection.** To generate YAP knockdown MDCK cells, MDCK cells were infected lentivirus containing shRNA targeting YAP and control plasmids gifted from Dr. Kun-Liang Guan's Lab (USCD). First, HEK293T cells were transfected with plasmids, vesicular stomatitis virus glycoprotein (VSVG) (Addgene, 8454), and D8.2 dvpr (Addgene, 8455) with FuGENE 6 (Roche, 11814443001) and the medium was changed after 12 hours. Supernatant collected at 72 and 96 hours after transfection was transferred to MDCK culture with polybrene (Sigma-Aldrich) at a final concentration of 5μg/mL. Cells were selected in culture medium containing 2 μg/mL of puromycin (Sigma-Aldrich).

**Small interfering (siRNA) RNA transfection.** siRNA transfection was performed with siRNA reagent system (Santa Cruz, sc-45064) following manufacturer's instructions. Briefly, cells were replated in six-well plates to 40% confluence in antibiotic-free 10% FBS-containing medium. For each transfection, 6 μl of WT1 siRNA (Santa Cruz, sc-36846) and unconjugated control siRNA-A (Santa Cruz, sc-37007) in 100 μl transfection medium (Santa Cruz, sc-36868) was mixed with 6 μl of transfection reagent (sc-29528) in another 100 μl transfection medium. Mixtures were incubated for 30 min at room temperature. Each well was washed with 2 ml of transfection medium once and then filled with prepared reagent mixture plus 0.8 ml transfection medium. After incubation overnight, an additional 1 ml of medium supplemented with 20% FBS with 2% antibiotics was added to each well for another 24 h. Transfection mixture was removed and replaced with normal growth medium on the following day. All experimental measurements were performed 24 h following replacement of the medium.

**Plasmid and transfection.** The plasmid of pcDNA-Flag-YAP1 was purchased through addgene (18881) and transfected into cells using Lipofectamine 3000 (Thermo Fischer) according to the manufacturer's instructions. The protein was assessed 24 h later by western blotting.

**Immunofluorescence Staining.** The samples on the nanofabricated coverslip were fixed with ice-cold 4% paraformaldehyde for 20 minutes, washed two times with phosphate-buffered saline (PBS) and permeablized with 0.1% Triton X-100 in PBS for 5 minutes. After washing with PBS, cultures were blocked by 10% goat-serum for 1 hour, and then incubated with primary antibody against vinculin (1:200, Sigma-Aldrich), E-cadherin antibody (1:200, Cell Signaling, 3195), phospho-myosin light chain (1:200, Cell signaling, 3674), YAP (1:100, Cell signaling, 4912 and Santa Cruz, sc-15407), active β-catenin (1:100, Millipore, 05-665), vimentin (1:100, Abcam), and WT1 (1:100, Santa Cruz, sc-192) for 3 hours in room temperature. After washing, with secondary antibodies and Alexa fluor 594 conjugated phalloidin

(1:40, Molecular Probes) and Hoescht (Invitrogen), cultures were incubated for 1 hour in room temperature. The slides were mounted with anti-fade reagent (SlowFade gold, Invitrogen) and taken by inverted microscope (Zeiss Axiovert 200M) with a X40 oil immersion objective (Zeiss, 1.6 NA).

**Immunoblot analysis.** Cells were homogenized in RIPA buffer (Thermo Scientific®) with protease inhibitor cocktail (Thermo Scientific), then centrifuged at 12,000 x g at 4 °C for 20 min and the supernatant was collected. Nuclear protein extraction was performed with Thermo Scientific Pierce NE-PER nuclear and cytoplasmic extraction reagents following the manufacturer's protocol (Thermo scientific, 78833). For Rac1 pull-down assay, we used Rac1 pull-down biochem assay kit (Cytoskeleton Inc., BK035) and followed the manufacturer's protocol. Protein concentration was quantified using BCA assay kit (Thermo scientific, 23227). Protein samples were subsequently diluted with sample buffer, heated at 70 °C for 10 minutes. These samples were separated on a 4-20% w/v SDS PAGE gel (BioRad), and transferred to a nitrocellulose membrane (BioRad). This membrane was blocked for 1 hour in blocking solution, TBST (10 mM Tris, pH 8.0 and 0.1% v/v Tween 20), supplemented with 5% BSA (BioRad) and incubated in 1:500 diluted solution of E-cadherin antibody (Cell Signaling, 3195), 1:1000 diluted solution of myosin light chain antibody (Abcam, ab48003), 1:500 diluted solution of phosphor-myosin light chain antibody (Cell Signaling, 3674), 1:1000 diluted solution of YAP antibody (Cell Signaling, 4912 and Santa Cruz, sc-15407), 1:1000 diluted solution of phospho-YAP antibody (Cell Signaling, 4911), 1:100 diluted solution of TRIO antibody (Santa Cruz, sc-28564), 1:1000 diluted solution of phospho-PAK (Cell Signaling, 2601), 1:1000 diluted solution of PAK (Cell Signaling, 2602), 1:1000 diluted solution of phospho-Merlin (Cell Signaling, 9163), 1:1000 diluted solution of Merlin (Cell Signaling, 6995), 1:1000 diluted solution of angiomotin (Santa Cruz, sc-98803), 1:1000 diluted solution of WT1 antibody (Santa Cruz, sc-192), 1:2000 diluted solution of laminB1 antibody (Abcam, ab28129) and 1:1000 diluted solution of GAPDH antibody (Abcam, ab9483) in blocking solution at 4 °C overnight. After washed (three times for 10 minutes) in TBST, goat anti-rabbit and anti-mouse secondary antibodies for Odyssey® western blotting (Li-cor) or HRP conjugated secondary antibodies (Pierce, 31160 (anti-mouse) and 31188 (anti-rabbit)) were treated. Finally, after washing with TBST, Odyssey® CLx infrared imaging system (Li-cor) or Bio-rad gel imaging system, ChemiDoc$^{TM}$ XRS+ (Bio-rad) with Pierce ECL Western Blotting Substrate (32106) was applied to blotting.

**Co-immunoprecipitation analysis.** Co-immunoprecipitation (co-IP) was performed using the Thermo Scientific Pierce co-IP kit following the manufacturer's protocol (Thermo scientific, 26140). Briefly, 60 μl of YAP antibody (Santa Cruz, sc-15407), 60 μl of angiomotin (Santa Cruz, sc-98803), and 30 μl of IgG antibody (Santa Cruz, sc-2027) was first immobilized for 2 hours using AminoLink Plus coupling resin. The resin was then washed and incubated with 200 μl (500 ug of proteins) of lysate harvested using

IP/lysis buffer in the kit and pre-cleaned with control agarose resine for 1 hour. After incubation, the coupling resin was again washed and protein eluted using elution buffer. Samples were analyzed by immunoblotting.

**Chromatin immunoprecipitation analysis.** Chromatin immunoprecipitation (ChIP) was performed using the Thermo Scientific Pierce agaross ChIP kit following the manufacturer's protocol (Thermo scientific, 26156). 1 x $10^6$ of cells per well was cross-linked with formaldehyde and their lysates were digested by 1U of MNase for 5 minutes in a 37°C water bath. Then, we immunoprecipitated digested-lysates with 5 μL of YAP (Santa Cruz, sc-15407X), WT1 (Santa Cruz, sc-192X), RNA polymerase II antibodies and 1μL of rabbit IgG. All DNA in digested lysates has an average length of 0.25 to 1 kb. PCR amplification was performed in 50 μL of mixture with Q5 high-fidelity DNA polymerase (NEB, M0491) and specific primers. The ~150-bp fragment of canine E-cadherin promoter was amplified with the primers 5′-CCCGCCGCAGGTGCAGCCGCAGC-3′ (direct) and 5′-GAGGCGGCGCGAGGCCGGCAG-3′ (reverse) [33]. PCR was carried out the following program, 25 cycles at 94°C for 40 s, 65 to 68 °C for 40 s, and 72 °C for 40 s. The amplified DNA was separated on 1 % agarose gel and visualized with ethidium bromide.

**Time-lapse microscopy.** Images were acquired after removing stencils, i.e., after allowing free expansion of cell sheets. The custom-made multi-well chamber integrated with the topographically patterned substratum was mounted onto the stage of a motorized inverted microscope (Zeiss Axiovert 200M) equipped with a Cascade 512B II CCD camera that has the environmental chamber containing. Phase-contrast images of cells were automatically recorded using the Slidebook 4.1 (Intelligent Imaging Innovations, Denver, CO) for 6 hours at 5 min intervals.

**Cell tracking.** We tracked cells manually in every image taken every 20 minutes in a time lapse imaging for 6 hours with the aid of a customized MATLAB (The MathWorks, Natick, MA) codes. The positions of cells with the respect to time after identifying and tracking individual every cell were used to analyze the migratory behavior. Statistical analyses were performed using unpaired two-sided Student's t-tests. Every experimental condition was analyzed in duplicates and at least 80 cells were tracked in each data replicate.

FIGURES LEGENDS

**Figure 1. Anisotropic texture of the mechanical cell environment increases EMT via YAP**
**a.** Schematic depiction of the edge areas of the expanding epithelial sheets, including formation of the Finger-Like Protrusions (FLPs) led by 'tip' cells and occasional separation of the tip cells ('dissemination'). The inset shown the electron micrograph of a profile of the NRA surface used in the experiments. **b.** Cell trajectories colored to reflect the cell speed values in an expanding sheet on flat (top) and textured NRA (bottom) substrata. The colors of each dot indicate the mean speed values of corresponding cells (left). The trajectories were tracked for 6 hours (right) with the speed measured every 20 minutes. **c.** Values of migration speed of individual cells on flat surface and nano-topographic substrata at different initial distances from the edge of the sheet (corresponding to values of $d$ in panel (a)). Each dot represents the average speed of an individual cell. Dashed lines indicate the averaged speed of isolated individual cells on flat surface (red) and NRA (blue). All error bars are S.E.M (# = Statistical significance of speeds on the marginal region vs. the most sub-marginal region of cells on flat surface (red) and NRA (blue), $^{\#\#}P<5\times10^{-4}$ and $^{\#\#\#}P<5\times10^{-6}$, * = Statistical significance of speeds of cells on flat surface vs. NRA, $*P<0.05$, $**P<5\times10^{-4}$, and $***P<5\times10^{-6}$, all two-sided Student's t-test). **d.** Numbers of disseminations from epithelial sheets per unit length of epithelial sheets on flat substrata and NRA for 18 hours (n ≥ 4). All error bars are S.E.M. (* = Statistical significance of numbers of dissemination on flat surfaces vs. NRA. $***P<5\times10^{-6}$, two-sided Student's t-test). **e.** The expression of Snail in the cells on flat and NRA surfaces analyzed by immunoblotting (n=3). All error bars are S.E.M. (* = Statistical significance of Snail expression in cells on flat vs. NRA substrata, $*P<0.05$, all two-sided Student's t-test). **f.** Dissemination of cells from epithelial sheets cultured on flat substrata and NRA in the presence of TGFβ (n ≥ 4). All error bars are S.E.M. (* = Statistical significance of the number of disseminations with control vs. drugs on the flat substratum (red) and on NRA (blue), flat substrata vs. NRA with the drug (black), $*P<0.05$, $**P<5\times10^{-4}$, and $***P<5\times10^{-6}$; all two-sided Student's t-tests). **g.** Cells on flat substrata and NRA analyzed by immunoblotting using YAP and phosphorylated YAP antibodies. **h.** Cell migration speeds of individual YAP$^{KD}$ cells in cell sheets as a function of the distance from the sheet edge on NRA. All error bars are S.E.M (# = Statistical significance of speeds on the marginal region vs. the most sub-marginal region of YAP$^{KD}$ cells on NRA, $^{\#\#\#}P<5\times10^{-6}$. * = Statistical significance of speeds of control vs. YAP$^{KD}$ cells on NRA (black) and of YAP$^{KD}$ cells on flat surface vs. NRA (green), $*P<0.05$, $**P<5\times10^{-4}$, and $***P<5\times10^{-6}$, all two-sided Student's t-test). **i.** Immunofluorescence staining for YAP and active β-catenin in epithelial cell sheets cultured on flat substrata and NRA. Translocation of YAP and active β-catenin into nuclei observed in marginal zones and FLPs of sheets expanding on NRA (brown boxes) and mostly cytoplasmic YAP and active β-catenin localization in sub-marginal cells on NRA (red boxes) and on flat substrata. **j.** Fractions of nuclei displaying different intensities of YAP and active β-catenin staining as a function of the distance from the sheet edge. All error bars are S.E.M.

**Figure 2. Cell-cell loosening effect of YAP and similar localization of WT1 with YAP in epithelial cell sheets on NRA**

**a.** Cell migration speeds of individual cells in cell sheets as a function of the distance from the sheet edge on flat substrata and NRA, in the presence of an E-cadherin functional blocking antibody. Dashed line with a square marker indicates the average cell migration speed of isolated control cells and solid line with a circle marker corresponds to that of isolated $YAP^{KD}$ cells. All error bars are S.E.M. (# = Statistical significance of speed values in the marginal region vs. the most sub-marginal region of $YAP^{KD}$ cells with drugs, $^{\#\#\#}P<5\times10^{-6}$. * = Statistical significance of control vs. $YAP^{KD}$ cells with drugs (green) and of $YAP^{KD}$ cells with vs. without drugs (black), $*P<0.05$, $**P<5\times10^{-4}$, and $***P<5\times10^{-6}$; all two-sided Student's t-test). **b.** Control, $YAP^{KD}$ and $YAP^{OE}$ cells were immunoblotted using active β-catenin, β-catenin and E-cadherin antibodies. **c.** Immunofluorescence staining for WT1 in epithelial cell sheets on a flat substratum and NRA. Translocation of WT1 into nuclei observed in marginal zones and FLPs of sheets expanding on NRA (brown boxes), sub-marginal cells on NRA (red boxes) and on flat substrata show YAP in cytoplasm. **d.** Control and $YAP^{KD}$ cells and their nuclear fractions were analyzed using immunoblotting using YAP and WT1 antibodies. **e.** Immunofluorescence staining for WT1 in cells and **f.** immunoblot assay of YAP and WT1 abundance in the nuclei in different density-epithelial cell sheet showing density-dependent translocation of WT1 into nucleus.

**Figure 3. YAP regulation of E-cadherin through WT1 in epithelial cell sheets on NRA**

**a.** Immunofluorescence staining for WT1 in $YAP^{KD}$ cell sheets (top), and for YAP in $WT1^{KD}$ cells cell sheets cultured on NRA. **b.** Co-IP analysis using YAP antibody, followed by immunoblotting using WT1 antibody. **c.** Chromatin immunoprecipitation (ChIP) analysis performed using YAP and WT1 antibodies, with IgG (negative control) and RNAPII and input genomic DNA (positive control), suggests WT1-YAP complex binding at E-cadherin promoter (n=3). All error-bars are S.E.M (* = Statistical significance of PCR products from each sample vs. IgG, $*P<0.05$ and $**P<0.01$; all two-sided Student's t-test). **d.** E-cadherin expressions of control and $WT1^{KD}$ cells analyzed by immunoblotting with WT1 and E-cadherin antibodies. All error-bars are S.E.M (n=3, * = Statistical significance of E-cadherin expression of control and $WT1^{KD}$ cells, $***P<0.005$; two-sided Student's t-test). **e.** Cell migration speeds of individual cells in control and $WT1^{KD}$ epithelial cell sheets as a function of the distance from sheet edge on NRA in the presence of an E-cadherin blocking antibody. All error bars are S.E.M. (# = Statistical significance of speeds in the marginal region vs. the most sub-marginal region of control (blue) and $WT1^{KD}$ (red) cells with E-cadherin blocking, n.s = non-significance and $^{\#\#}P<5\times10^{-3}$. * = Statistical significance of control vs. WT1$^{KD}$ cells with drugs (black) and of WT1$^{KD}$ cells with vs. without treatment (red), n.s = non-significance, *$P<0.05$, **$P<5\times10^{-4}$ and ***$P<5\times10^{-6}$, all two-sided Student's t-test.) **f.** Dissemination of cells in control and WT1$^{KD}$ epithelial sheets on NRA in the presence of an E-cadherin blocking antibody (n ≥ 3). All error bars are S.E.M. (* = Statistical significance of dissemination of control cells (blue) and of WT1$^{KD}$ cells (red) with vs. without treatment, **$P<0.01$, and n.s = no statistical significance, all two-sided Student's t-tests). **g.** Immunoblotting of phosphorylated YAP and total YAP in cell sheets cultured in the presence of different concentrations of E-cadherin blocking antibody. All error bars are S.E.M (n=3). **h.** Schematic of the YAP-mediated cell dissemination triggered by mechanical cues stemming from NRA through WT1 and E-cadherin leading to cell dissemination in epithelial cell sheets on NRA.

**Figure 4. Cross-regulation of YAP, Rac1 and Merlin in epithelial cell sheets cultured on NRA**

**a.** Rac1 activity assay using cells cultured on flat and NRA surfaces (n=4). All error bars are S.E.M. (* = Statistical significance of the ratio of active Rac1 to Rac1 on flat surfaces vs. NRA. *$P<0.05$, two-sided Student's t-test). **b**. Control, YAP$^{KD}$ and YAP$^{OE}$ cells were immunoblotted to evaluate Rac1 activity via pull-down assay, expression of TRIO and phosphorylation of PAK and Merlin. **c.** Cell migration speeds of individual cells in YAP$^{KD}$ (green, left) and YAP$^{OE}$ epithelial cell sheets (purple, right) as a function of the distance from the sheet edge on NRA in the presence of Rac1 inhibitor, NSC23766 (top) and a TRIO inhibitor, ITX3 (bottom). All error bars are S.E.M. (# = Statistical significance of speed values in the marginal region vs. the most sub-marginal region of YAP$^{KD}$ cells, $^{\#}P<5\times10^{-2}$, $^{\#\#}P<5\times10^{-4}$, and $^{\#\#\#}P<5\times10^{-6}$. * = Statistical significance of control vs. YAP$^{KD}$ cells with drugs (green), control vs. YAP$^{OE}$ cells with drugs (purple) and of YAP$^{KD}$ /YAP$^{OE}$ cells with vs. without drugs (black), *$P<0.05$, **$P<5\times10^{-4}$, and ***$P<5\times10^{-6}$, all two-sided Student's t-test.) **d.** Immunofluorescence staining for Merlin in epithelial cell sheets cultured on a flat surface and NRA. In the marginal zone and FLPs of sheets on NRA, Merlin was primarily localized in cytosol (brown boxes), whereas it was localized in the cell-cell contacts of sub-marginal cells cultured on NRA and flat substrata (red boxes). **e.** Cells on flat and NRA surfaces were used for Co-IP analysis with AMOT antibody for investigating physical bounding between AMOT and a target protein, Merlin, and the precipitates were immunoblotted using Merlin antibody. **f.** Lysates from control and YAP$^{KD}$ cells were used for co-imunoprecipitation (Co-IP) study using the AMOT antibody, analyzed by immunoblotting with Merlin antibody. **g.** Control and Merlin$^{KD}$ cells were immunoblotted for evaluation of Rac1 activity via pull-down assay, YAP expression and its phosphorylation. All error bars are S.E.M. (n=3, * = Statistical significance of the ratio of active Rac1 to Rac1 on control vs. Merlin$^{KD}$ cells. ***$P<5\times10^{-3}$, two-sided Student's t-test). **h.** Cell migration speeds of individual cells in Merlin$^{KD}$ epithelial cell sheets as a function of the distance from sheet edge in cell sheets cultured on NRA in the presence of a Rac1 inhibitor, NSC23766. All error bars are S.E.M. (# = Statistical significance of speed values in the marginal region vs. the most sub-marginal region of Merlin$^{KD}$ cells, $^{\#\#\#}$P<5x10$^{-6}$. * = Statistical significance of control vs. Merlin$^{KD}$ cells without drugs (blue) and of Merlin$^{KD}$ cells with vs. without drugs (black), *P<0.05, and **P<5x10$^{-3}$, all two-sided Student's t-test.) **i.** Dissemination of cells in control and Merlin$^{KD}$ epithelial sheets on NRA (n ≥ 3). All error bars are S.E.M. (* = Statistical significance of the number of disseminations of cells from control vs. Merlin$^{KD}$ epithelial sheets, *P<0.05, all two-sided Student's t-test). **j.** Schematic of YAP-mediated signaling cascades via Rac1 and Merlin causing EMT and cell dissemination from epithelial sheets on NRA.

**Figure 5. Validation of the model composed of double negative feedback (YAP- E-cadherin) or double positive feedback (YAP-Rac1) explaining EMT on NRA a.** Simulation result exhibiting two stable states of high and low YAP activity indicating epithelial (low Rac1, high E-cadherin and low YAP) and mesenchymal (high Rac1, low E-cadherin and high YAP) states depending on Rac1 and E-cadherin activity in this system. **b.** Cell migration speed on NRAs with different rigidity values suggesting that rigidity can regulate the basal rate of YAP activation at different initial distances from the edge of the sheet. All error bars are S.E.M (* = statistical significance of speeds of cells on NRA which rigidities are 10 MPa vs. 1 GPa, *P<5x10$^{-2}$, **P<1x10$^{-2}$ and ***P<5x10$^{-3}$, all two-sided Student's t-test). **c.** Simulated bi-modal distribution of YAP activity as a function of the distance from sheet edge on NRA substrata of different rigidity values **d.** Rigidity dependent YAP localization in nuclei of cells cultured on NRA. Immunofluorescence staining of YAP showing nuclear localization at different distances from the sheet edge on NRAs with different rigidity values (top). Fractions of nuclei displaying different intensities of YAP staining as a function of the distance from the sheet edge on NRA having different rigidity (bottom) (see details in Supplemental discussion and Supplemental figure 8)

**Figure 6. The clinical relevance of EMT-like cell behavior elicited by mechanical cues to renal cancer**
**a.** Difference of signaling of renal cell carcinoma cell, ACHN, related to EMT, cell-cell adhesion, cell migration between disseminated cells and cells remaining within monolayer after dissemination elicited by mechanical cues. **b.** Analysis of patient data, sequenced tumors (451 samples) of kidney renal clear cell carcinoma, (TCGA, provisional) via cbioportal (www.cbioportal.org). Oncoprints show alterations in YAP1, and WT1 across a set of kidney renal clear cell carcinoma. Z scores for mRNA expression (RNA Seq V2RSEM) and protein expression were 2.326 for 99% confidentiality (one-tail). **c.** Disease free survival Kaplan-Meier Estimate of altered YAP1 (left) and YAP1/WT1 (right). P-values were calculated with Fisher exact test.

Acknowledgements

We thank Dr. Kun-Liang Guan for sharing shRNA plasmids; Patrick Conlon for expert assistance; Alex Rhee for sharing reagents; Sung Hoon Lee for fabricating 3D mold for microstencil. J.P is a recipient of Samsung scholarship. This work was also supported by NIH grants R01NS070024 to (AL and AQ-H), and U54CA209992 and U01CA155758 (AL)

Author contributions

J.P* and D. K* contributed equally. J.P and D. K conceived and designed the project. J.P, S.R.S and P. K performed the experiments. H.K and D. K designed and fabricated substrata with the supervision of K.Y.S. J.P and P.K tracked and analyzed migration time-lapse data. K created YAP$^{KD}$ and control cell lines. D.E designed and created microstencils. J.P and S.R.S. interpreted the results. J.P simulated computational model. A.L. and A.Q.-H. supervised the project. J.P and A.L. wrote the manuscript. A.L. and A.Q.-H. reviewed and revised the manuscript.

Author information

The authors declare no competing financial interests. Correspondence and requests for material should be addressed to A.Q.-H (Quinones-Hinojosa.Alfredo@mayo.edu) or A.L. (andre.levchenko@yale.edu).

SUPPLEMENTAL FIGURE LEGENDS

**Figure S1. Procedure of making collective cell sheets on NRA**. **a.** Microstencil of PDMS. NRA surfaces are coated with 30 μg/mL of collagen type I for 3 hours. After aspirating the solution of collagen type I, microstencils defining the size of the initial epithelial sheets are bonded to the nano-topographically defined substratum. In each rectangular hole of a microstencil, 2.0 x 10$^5$ cells diluted in 50 μL of media are seeded. After full cell attachment to the substratum overnight, the microstencil is peeled off. Thereafter, the expansion of epithelial cell sheets is tracked using microscopy. **b.** Collective cell migration for 16 hours on flat and NRA substrata in a single dish. **c.** Epithelial sheet expansion perpendicular to the direction of the nano-ridge array of NRA for 18 hours, showing a lack of FLPs and a lower degree of expansion vs. that in the direction of the ridges.

**Figure S2. Increases in the formation of FLP on NRA. a.** An example of FLPs formed on an NRA. **b.** Immunofluorescent staining for actin of cells on a flat and anisotropic surfaces (left). An arrow marks lack of cortical actin at the tip of FLPs and an asterisk points out concentrated actin at the convex region of the edge. Double-sided arrows indicate the direction of groove/ridge arrays. Phosphorylated myosin

light chain (pMLC) at the frontal periphery of FLPs on a flat surface and NRA (right). An arrow indicates weak pMLC at the tip of a FLP on NRA and asterisks point at concentrated pMLC at the sides of the FLP. Double-sided arrows indicate the direction of groove/ridge arrays.

**Figure S3. Focal adhesion and cytoskeleton of epithelial cell layers on NRA. a.** Images of cells immunofluorscently stained for vinculin on a flat surface and NRA **b.** Captured live-cell imaging of GFP-F-actin transfected cells in the marginal and sub-marginal region of an epithelial sheet. **c.** Captured live-cell imaging of GFP-α-tubulin transfected cells at the tip and the convex region in the boundary of an epithelial cell sheet

**Figure S4. Molecular characterization of mechanically induced EMT. a.** Example of disseminated cells emerging from the advancing cell sheet cultured on NRA for 18 hours. Asterisks (*) indicate the disseminated cells **b.** Immunofluorescence staining for Slug and **c.** Twist expression in FLPs in cells cultured on flat and NRA surfaces. Asterisks (*) marks exclusive expression of EMT markers on FLPs on NRA **d.** Vimentin staining proximal to the edge of a cell sheet on a flat substratum and NRA.

**Figure S5. Suppressed EMT characteristics in $YAP^{KD}$ cells on NRA.** Decreased localization of β-catenin in nuclei (top) and expression of vimentin (bottom) in the marginal regions of $YAP^{KD}$ cell sheets cultured on NRA.

**Figure S6. YAP regulation of RhoA small GTPase in epithelial cell sheets cultured on NRA. a.** Immunofluorescent staining of pMLC of FLPs in control and $YAP^{KD}$ cells on NRA. The knockdown of YAP partially restores pMLC in punctate patterns at the tips of FLPs (marked by *). Double-side arrows indicate the direction of groove/ridge arrays. **b.** Control and $YAP^{KD}$ cells were immunoblotted for evaluating myosin light chain kinase activity using MLC and pMLC antibodies **c.** Cell migration speeds of individual cells in $YAP^{KD}$ epithelial cell sheets as a function of the distance from sheet edge on NRA with ROCK inhibitor, Y27632. All error bars are S.E.M. (# = Statistical significance of speeds on the marginal region vs. the most sub-marginal region of ROCK inhibited $YAP^{KD}$ cells, $^{\#\#\#}P<5\times10^{-6}$. * = Statistical significance of control vs. $YAP^{KD}$ cells with drugs (green) and of $YAP^{KD}$ cells with vs. without drugs (black), $*P<0.05$, and $***P<5\times10^{-6}$, all two-sided Student's t-test.)

**Figure S7. Collective cell migration on NRA having different rigidity (10 MPa, 0.1 GPa and 1 GPa) regulated by YAP. a.** Immunofluorescence staining for E-cadherin and vinculin in epithelial cell sheets cultured on NRA of different rigidity displaying more aligned/elongated focal adhesions in FLPs on stiffer NRAs. **b.** Average intensity of YAP staining in nuclei on NRA substrata of different rigidity as a function of the distance from sheet edge. All error bars are S.E.M.

**Figure S8. Simulation of YAP-regulated signaling networks containing double-negative feedback with E-cadherin and positive feedback with Rac1. a.** Simulated YAP activity as a function of distance from sheet edge for different assumed rigidity values, represented by different basal activation of YAP (see the model description in the Supplemental discussion). A shaded region corresponds to the regions of epithelial sheets where we experimentally observed nuclear localization of YAP. **b.** Two-parameter bifurcation diagram relating E-cadherin- or Rac1-dependent rates of YAP activation and basal rate of YAP activation for different NRA rigidity values. This diagram shows the range of parameters where the system has bi-stability. Dashed lines indicate basal rates of YAP activation corresponding to assumed substrate rigidity values of NRA. Double-sided arrows indicate the range of E-cadherin- or Rac1-dependent rate of YAP activation where the output exhibits bi-stablity for the specific rigidity values.

**Figure S9. Raw images of immunoblotting experiments**

### SUPPLEMENTAL DISCUSSION

#### 1. Computational modeling of YAP driven signaling network regulating EMT

The YAP-driven signaling networks determining EMT and cell dissemination from epithelial sheets on NRA were modeled as a combination of reactions suppressing cell-cell adhesion through down-regulation of E-cadherin complexes and enhancing individual cell migration via small GTPases, particularly Rac1. The inhibitory effects of E-cadherin and stimulatory effect of Rac1 on YAP activation identified in our study suggest feedback interactions. The corresponding signaling processes summarized in Fig. 3h and Fig. 4j were modeled as a series of ordinary differential equations below. As many details of these processes remain to be elucidated, the model equations are simplified versions of the interactions capturing the their feedback character:

$$\frac{d}{dt}[\text{YAP}_{\text{E-cadherin}}] = k_{\text{YAP}} + k_{\text{YAP·E-cadherin}}[\text{E-cadherin}] - d_{\text{YAP}}[\text{YAP}_{\text{E-cadherin}}] \quad \text{(Fig. 3h)}$$

$$\frac{d}{dt}[\text{E-cadherin}] = k_{\text{E-cadherin}} \frac{K^3_{\text{E-cadherin}}}{K^3_{\text{E-cadherin}} + [\text{YAP}_{\text{E-cadherin}}]^3} - d_{\text{E-cadherin}}[\text{E-cadherin}]$$

$$\frac{d}{dt}[\text{YAP}_{\text{Rac1}}] = k_{\text{YAP}} + k_{\text{YAP·Rac1}}[\text{Rac1}] - d_{\text{YAP}}[\text{YAP}_{\text{Rac1}}] \quad \text{(Fig. 4j)}$$

$$\frac{d}{dt}[\text{Rac1}] = k_{\text{Rac1}} \frac{K^3_{\text{Rac1}}}{K^3_{\text{Rac1}} + [\text{YAP}_{\text{Rac1}}]^3} - d_{\text{Rac1}}[\text{Rac1}]$$

A major assumption underlying the model is that the dependence of E-cadherin and Rac1 on YAP activity are non-linear and saturable. These assumptions are necessary to obtain switch-like, bi-stable behavior. At the steady state, all the time derivatives are equal to zero, leading to the following algebraic equations:

$$0 = k_{YAP} + k_{YAP \cdot E\text{-cadherin}}[\text{E-cadherin}] - d_{YAP}[\text{YAP}_{E\text{-cadherin}}]$$

$$0 = k_{E\text{-cadherin}} \frac{K_{E\text{-cadherin}}^3}{K_{E\text{-cadherin}}^3 + [\text{YAP}_{E\text{-cadherin}}]^3} - d_{E\text{-cadherin}}[\text{E-cadherin}] \quad \text{(Fig. 3h)}$$

$$0 = k_{YAP} + k_{YAP \cdot Rac1}[\text{Rac1}] - d_{YAP}[\text{YAP}_{Rac1}]$$

$$0 = k_{Rac1} \frac{K_{Rac1}^3}{K_{Rac1}^3 + [\text{YAP}_{Rac1}]^3} - d_{Rac1}[\text{Rac1}] \quad \text{(Fig. 4j)}$$

These equations with parameters and their assumed values, listed in Supplemental table 1, constitute a full description of YAP-driven signaling network. In particular, we note the negative value of $k_{YAP\text{-E-cadherin}}$ represents a negative effect of E-cadherin on YAP in the double negative feedback determining cell-cell adhesion, $k_{E\text{-cadherin}} > 0$ represents negative regulation of YAP on E-cadherin resulting from binding of YAP-WT1 transcription complex to the E-cadherin promoter. Similarly, we take $k_{YAP\text{-Rac1}} > 0$, which indicates a positive effect of Rac1 on YAP, whereas the negative value of $k_{Rac1}$ accounts for the corresponding positive effect of YAP on Rac1 through facilitation of dissociation of Merlin from AMOT in the positive feedback regulating cell migration speed. The parameters are arbitrary, but the results are meant to be semi-quantitative in nature, to demonstrate the effect of the feedback interaction. Overall, the outcome of the model is qualitatively robust to these values, as demonstrated by the bifurcation diagram in Supplemental figure 8.

Each pair of equations generates two 'null clines' whose intersections are the steady states of the system. Given the model non-linearity, three steady states are robustly obtained, one unstable and two stable ones. This is the common 'bi-stability' regime, suggesting two mutually exclusive states in a switch-like regulation: if E-cadherin is low and Rac1 is high, YAP is high. Conversely, if Rac1 is low and E-cadherin is high, YAP is low (Fig. 5a). We suggest that these two states represent a mesenchymal state (low E-cadherin, high Rac1 and high YAP) and epithelial state (high E-cadherin, low Rac1 and low YAP) in epithelial-mesenchymal transition. This is also consistent with the bimodal distribution of EMT markers such as nuclear translocation of active β-catenin and YAP, as a function of the distance from sheet edge, as shown in Fig 1j. This behavior is robust to changes within the range of the parameter values, E-cadherin- or Rac1-dependent rates of YAP activation and the basal rate of YAP activation as long as the null clines continue to intersect in 3 distinct points, preserving the bi-stable nature of the output (see the bifurcation analysis in Supplemental figure 8). This model predicts that the range of where the system has bimodal distribution of EMT markers becomes wider as the basal rate of YAP activation increases. This is an important model prediction, validated in Fig. 5 of the main text. In this figure, we observed the extension of bi-stable region with increasing substratum rigidity, the input that we assume to correspond to higher values of basal rate of YAP activation (Fig. 5c and Supplemental figure 8). Basal rates of YAP activation for each NRA rigidity value are listed in Supplemental table 2.

**Supplemental table 1**

| Parameter | unit | Value | Description |
|---|---|---|---|
| $k_{YAP}$ | µM s$^{-1}$ | 1 | Basal rate of YAP activation |
| $k_{YAP\text{-}E\text{-}cadherin}$ | s$^{-1}$ | -1.8 | E-cadherin-dependent rate of YAP activation |
| $k_{YAP\text{-}Rac1}$ | s$^{-1}$ | 1.8 | Rac1-dependent rate of YAP activation |
| $k_{E\text{-}cadherin}$ | s$^{-1}$ | 0.9 | YAP-dependent rate of E-cadherin expression |
| $k_{Rac1}$ | s$^{-1}$ | -0.9 | YAP-dependent rate of Rac1 activation |
| $K_{E\text{-}cadherin}$ | µM | 1 | Dissociation constant of YAP-WT1 transcriptional constant |
| $K_{Rac1}$ | µM | 1 | Michaelis-like constant of Rac activation |
| $d_{YAP}$ | s$^{-1}$ | 1 | Inactivation rate of YAP |
| $d_{E\text{-}cadherin}$ | s$^{-1}$ | 1 | Degradation rate of E-cadherin |
| $d_{Rac1}$ | s$^{-1}$ | 1 | Inactivation rate of Rac1 |

**Supplemental table 2**

| Parameter | unit | Value | Description |
|---|---|---|---|
| $k_{YAP,\ 10\ MPa}$ | µM s$^{-1}$ | 0.95 | Basal rate of YAP activation on 10 MPa NRA |
| $k_{YAP,\ 0.1\ GPa}$ | µM s$^{-1}$ | 1 | Basal rate of YAP activation on 0.1 GPa NRA |
| $k_{YAP,\ 1\ GPa}$ | µM s$^{-1}$ | 1.05 | Basal rate of YAP activation on 1 GPa NRA |

**Supplemental video 1.** Collective migration of cells in an epithelial cell sheet on a flat substratum and NRA

**Supplemental video 2.** Actin of marginal and sub-marginal cells in an epithelial cell sheet on NRA

**Supplemental video 3.** α-microtubule of cells at the tip and convex region in the boundary of an epithelial cell sheet on NRA

**Supplemental video 4.** Dissemination of 'tip' cells on NRA with EMT inducer, TGFβ

**Supplemental video 5.** Collective migration of control and YAP$^{KD}$ cells on NRA

**Supplemental video 6.** Collective migration of control and YAP$^{KD}$ cells on NRA with E-cadherin inhibition

**Supplemental video 7.** Collective migration of control and WT1$^{KD}$ cells on NRA with E-cadherin inhibition

**Supplemental video 8.** Collective migration of control and YAP$^{KD}$ cells on NRA with Rac1 and TRIO inhibition

**Supplemental video 9.** Collective migration of control and YAP$^{OE}$ cells on NRA with Rac1 and TRIO inhibition

**Supplemental video 10.** Collective migration of control and Merlin$^{KD}$ cells on NRA with Rac1 inhibition

**Supplemental video 11.** Collective migration of control and YAP$^{KD}$ cells on NRA with ROCK inhibition

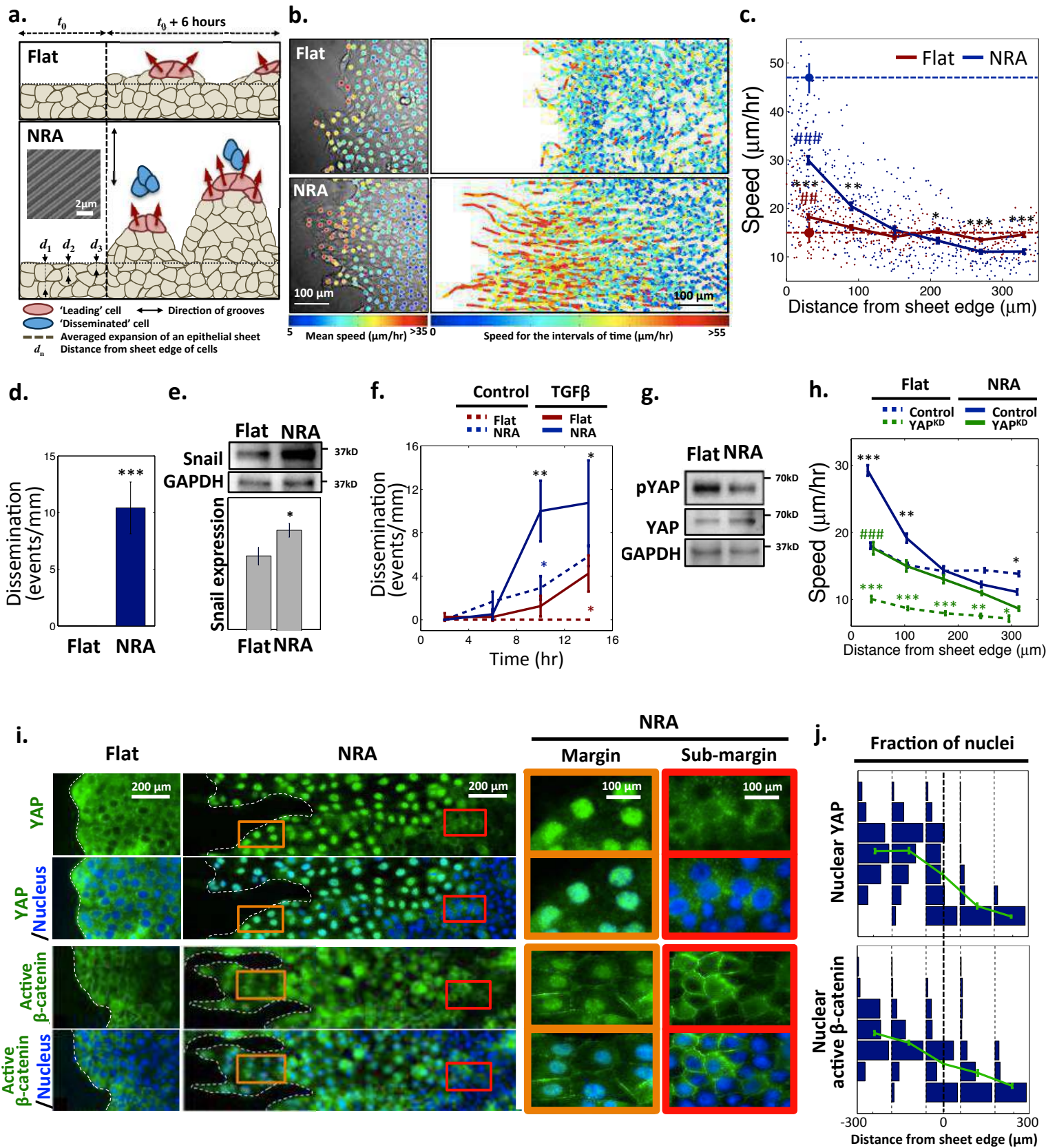

Figure 1

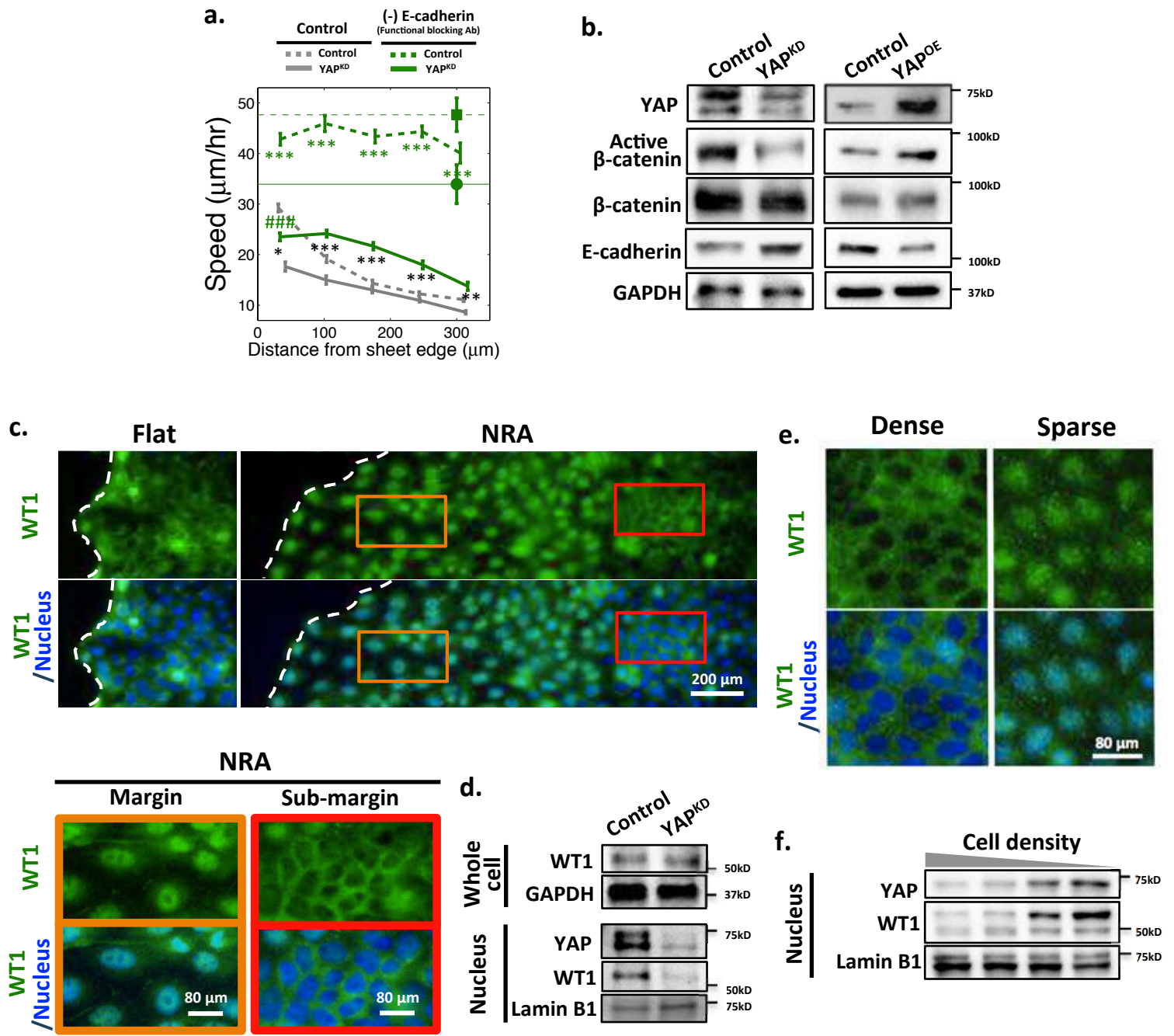

Figure 2

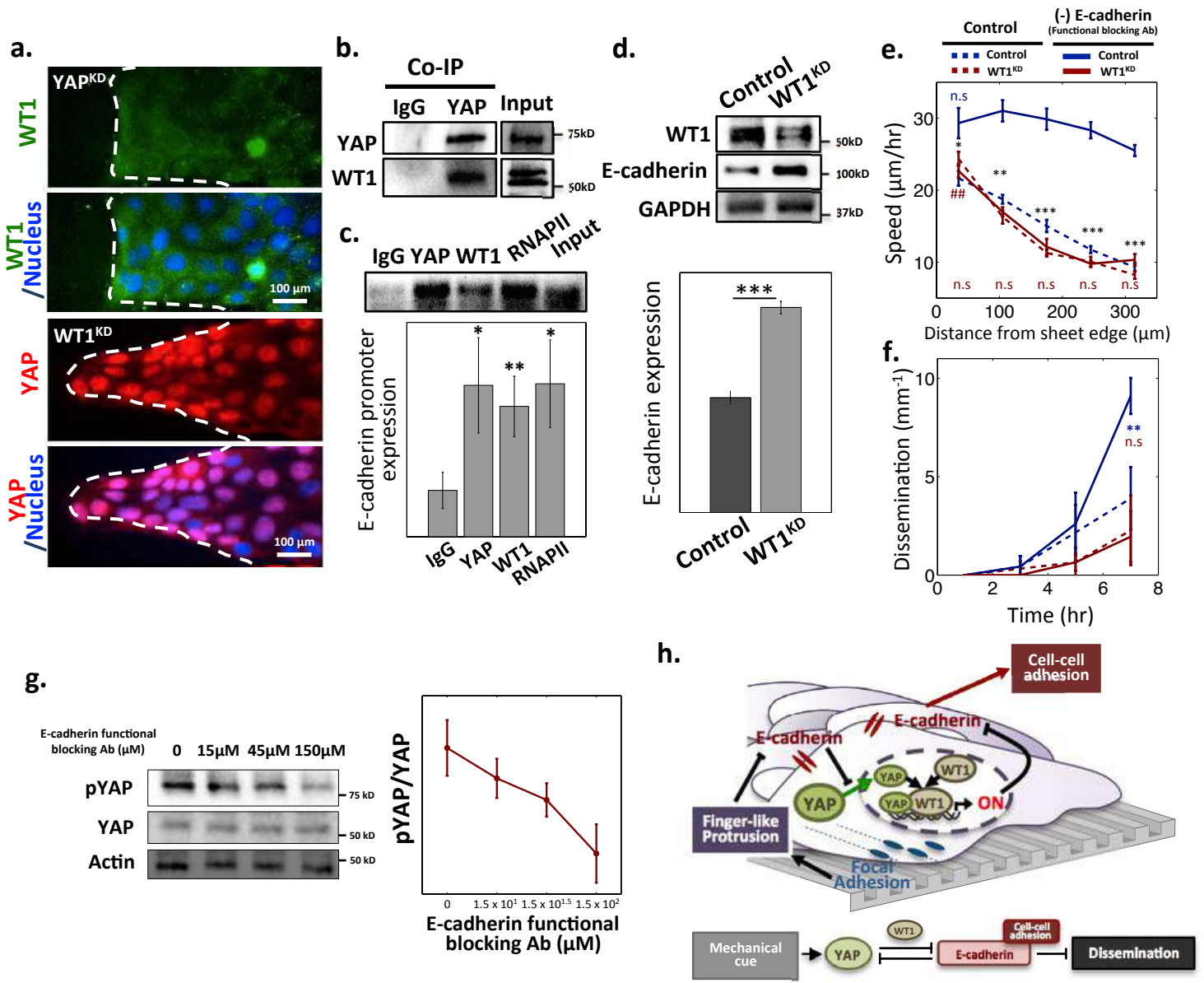

Figure 3

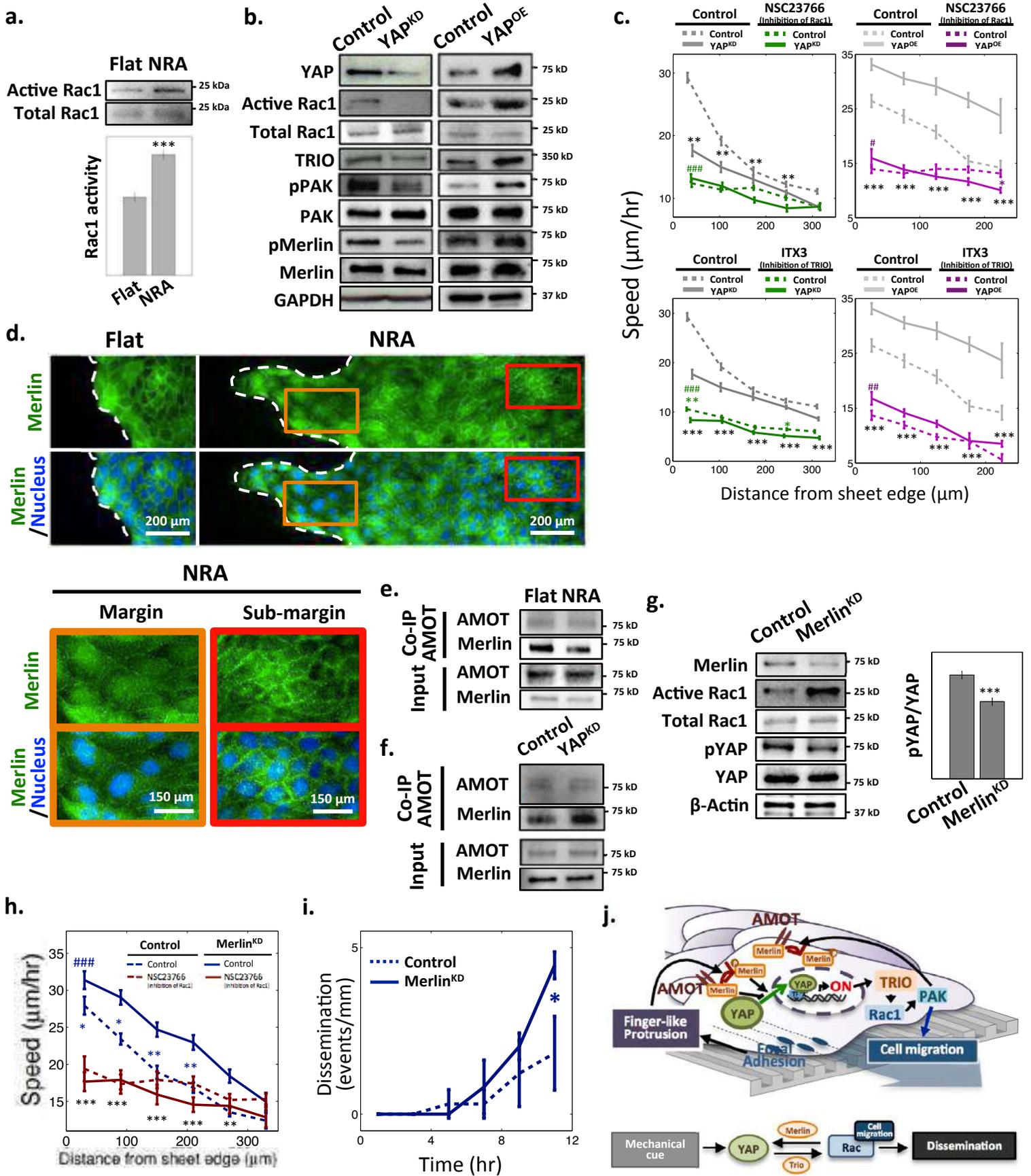

Figure 4

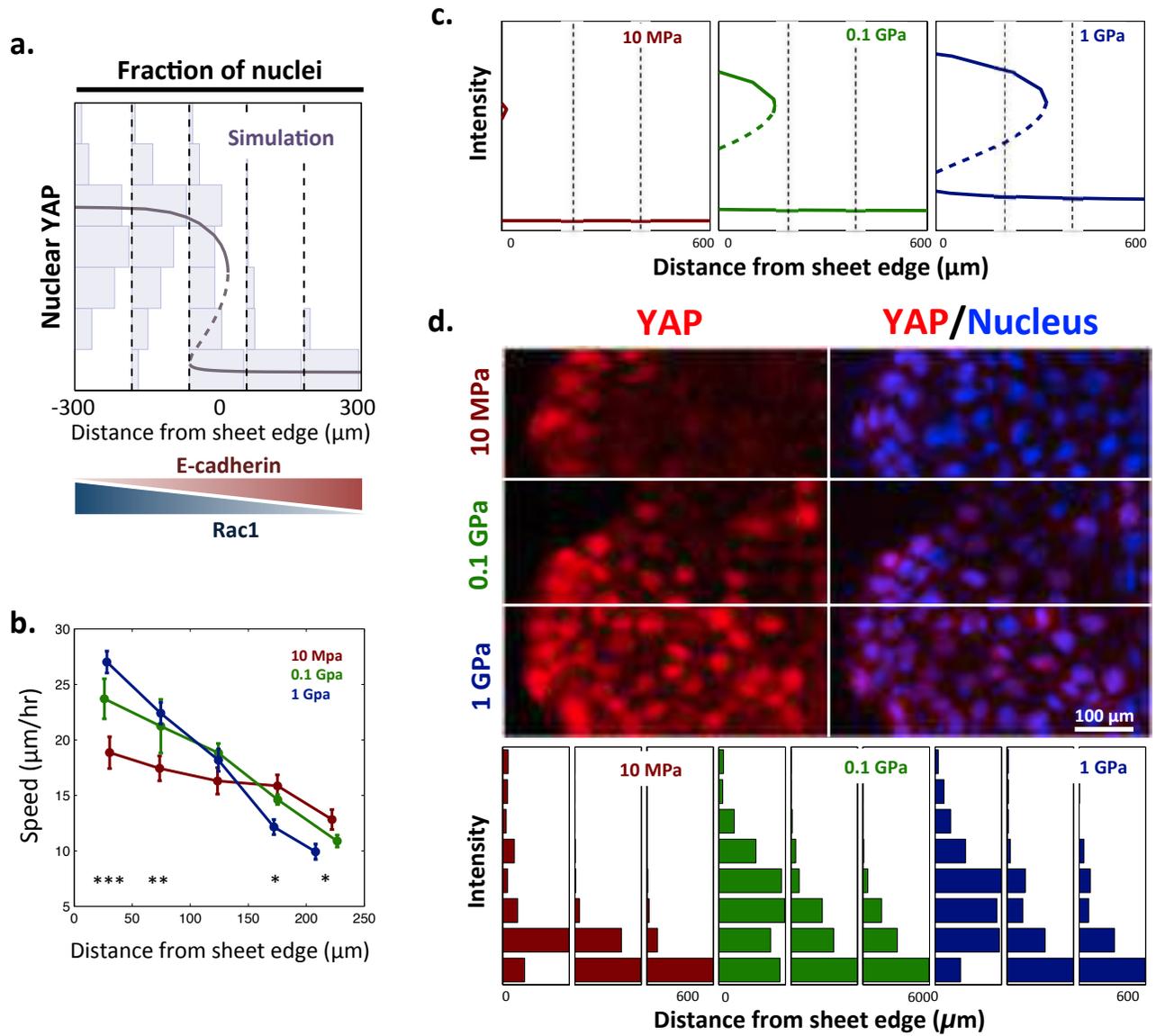

Figure 5

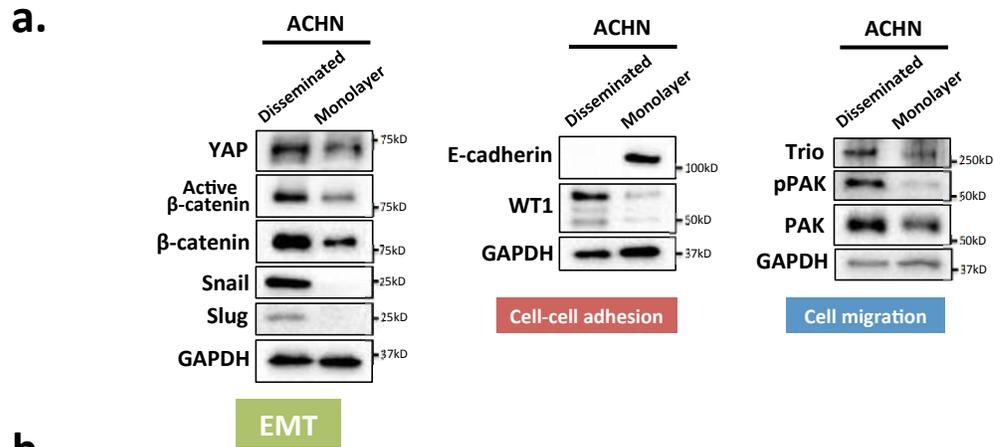

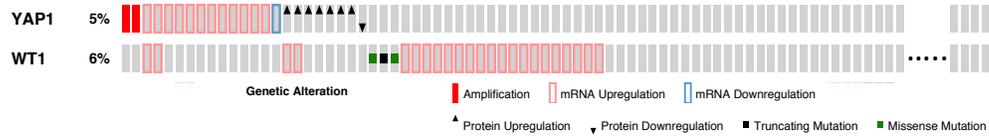

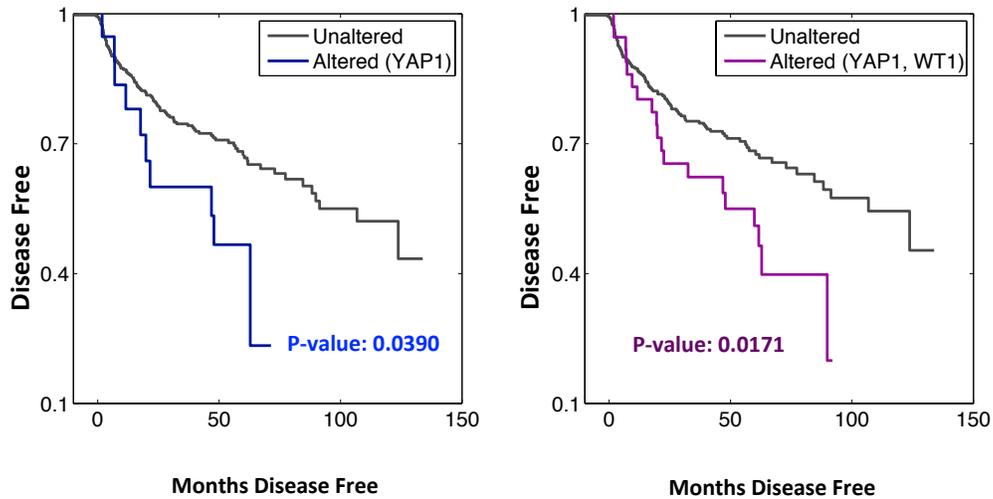

Figure 6

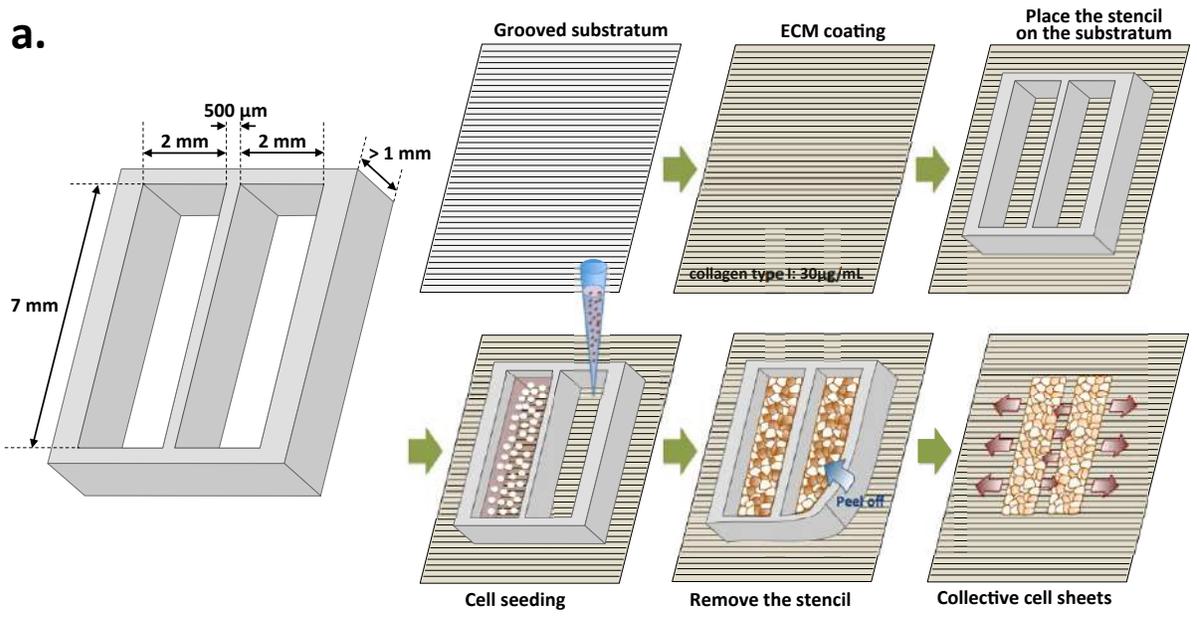

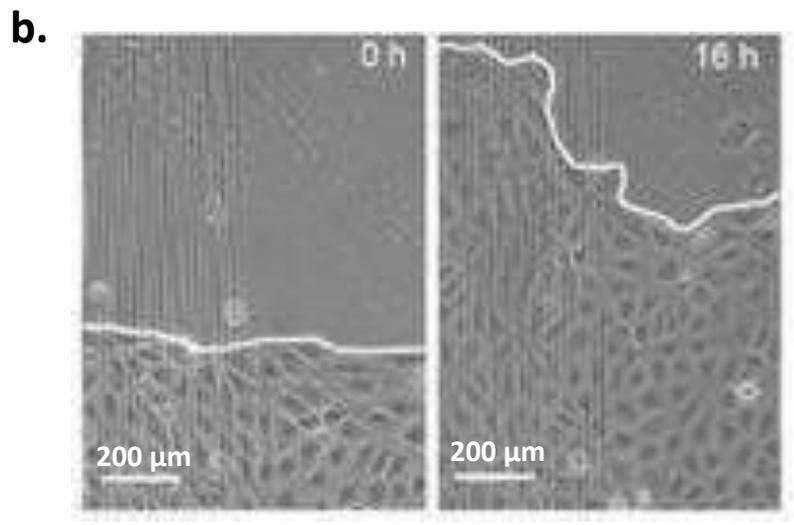

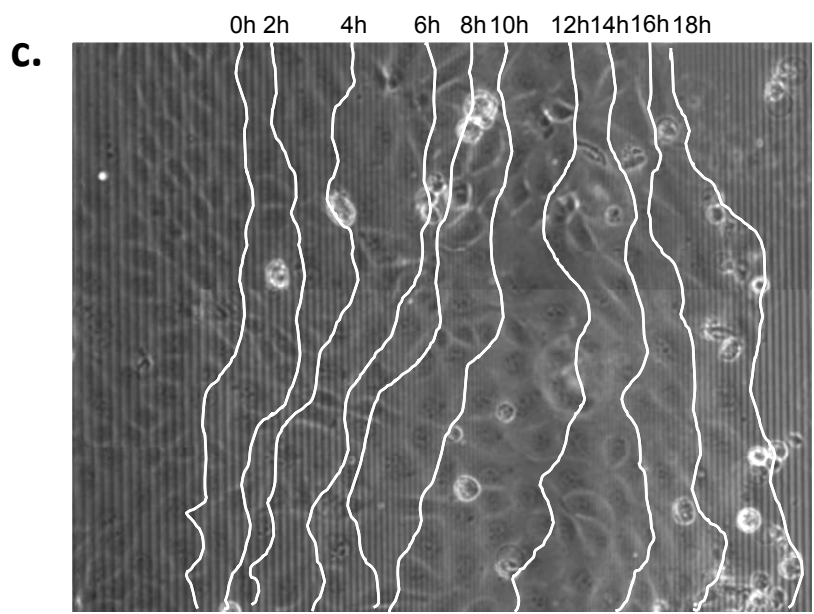

**Supplemental figure 1**

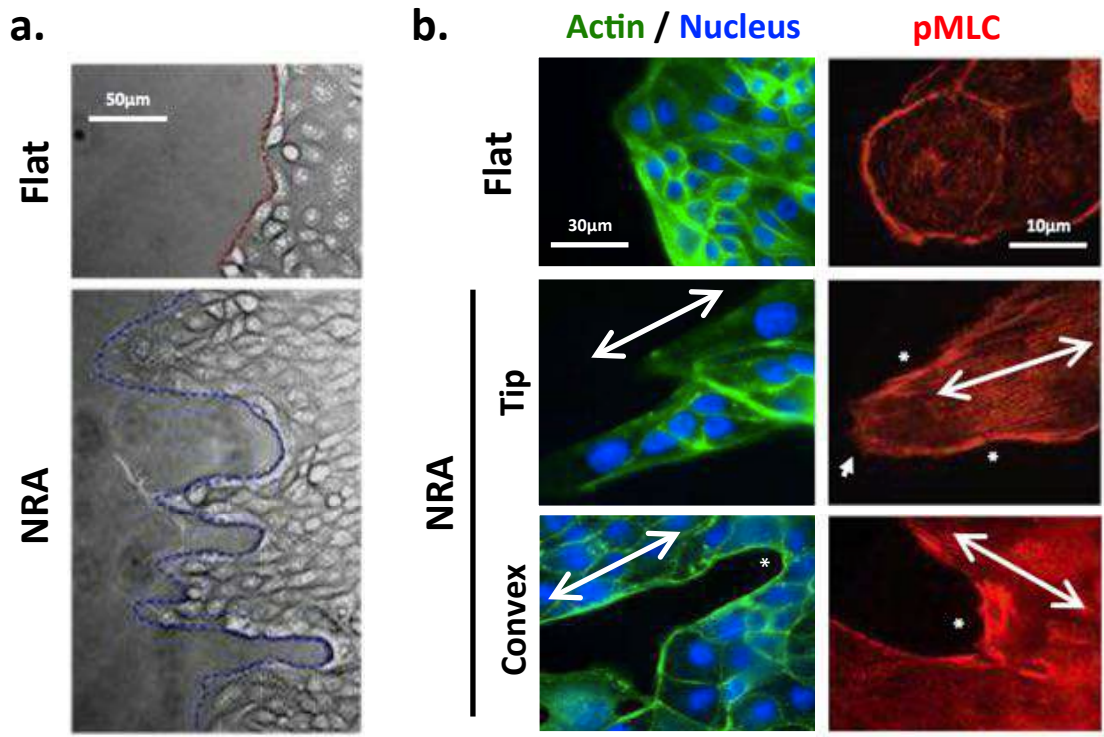

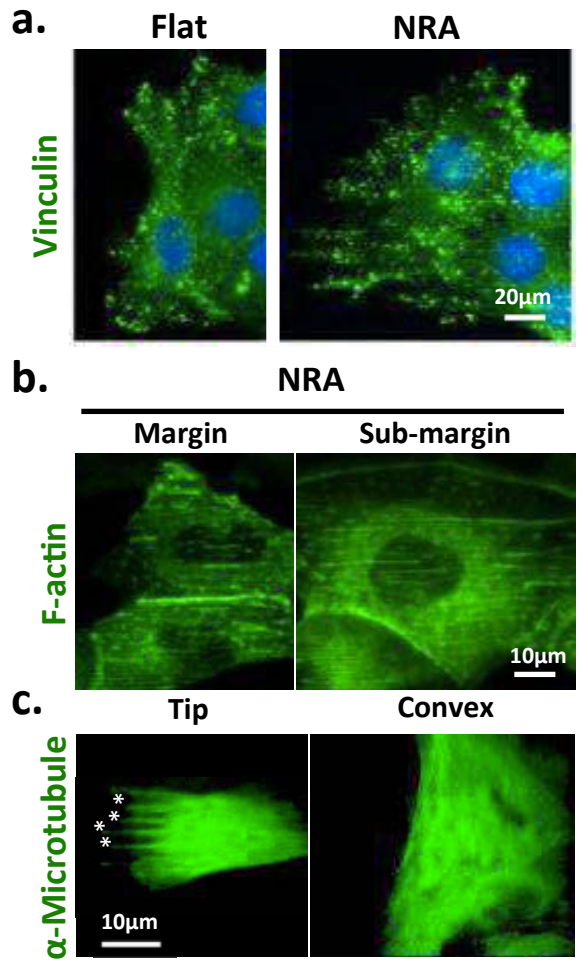

Supplemental figure 2

Supplemental figure 3

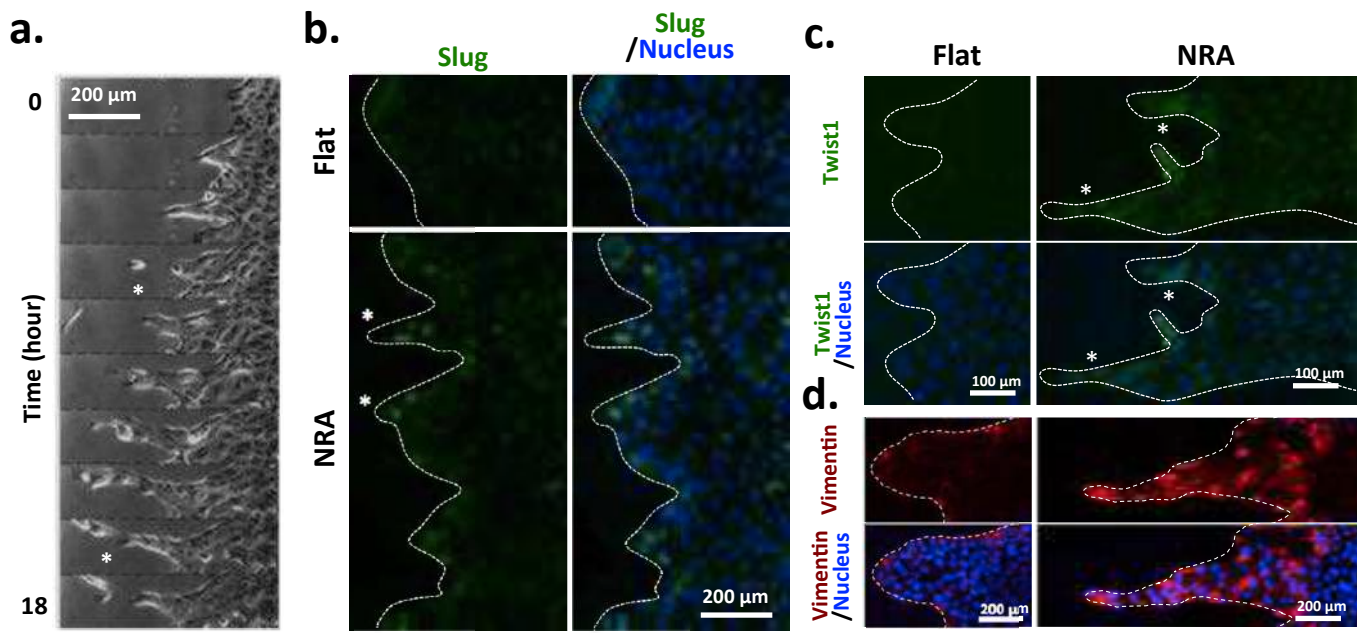

**Supplemental figure 4**

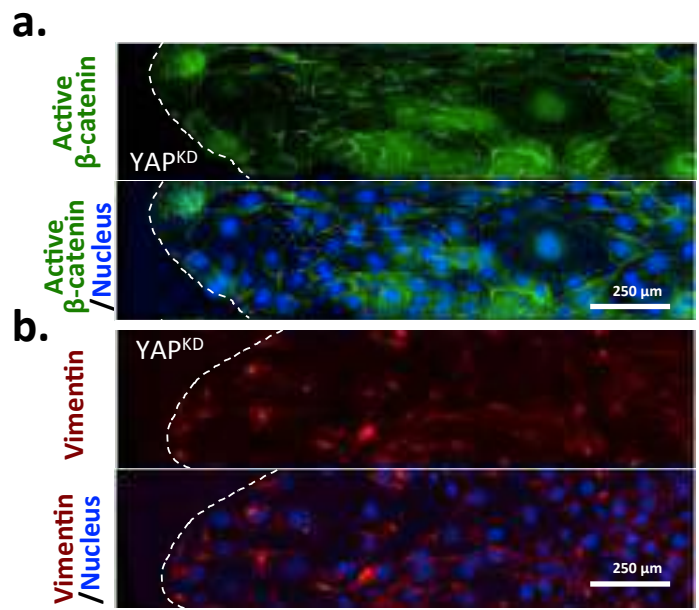

**Supplemental figure 5**

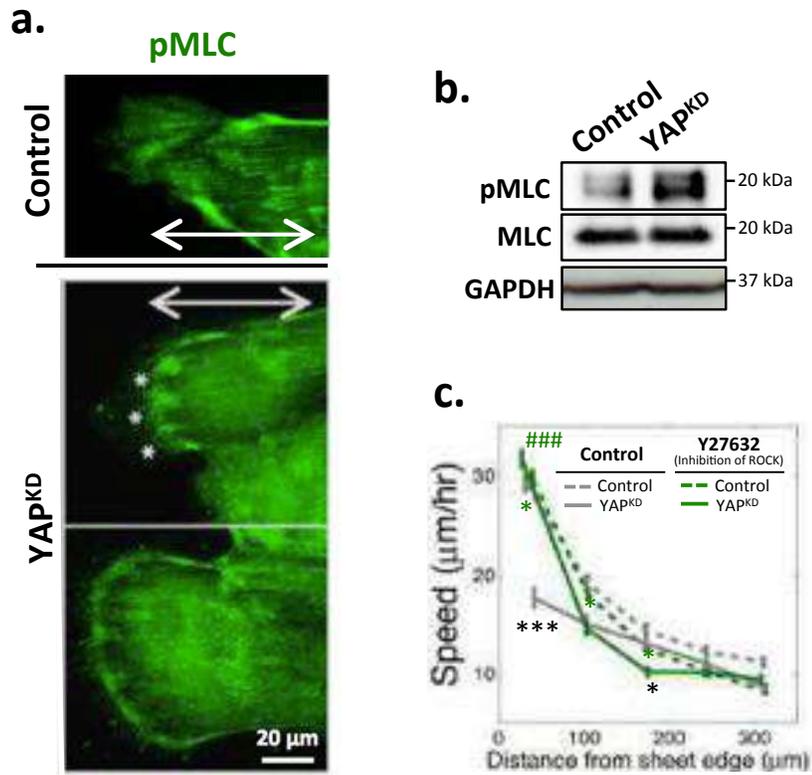

**Supplemental figure 6**

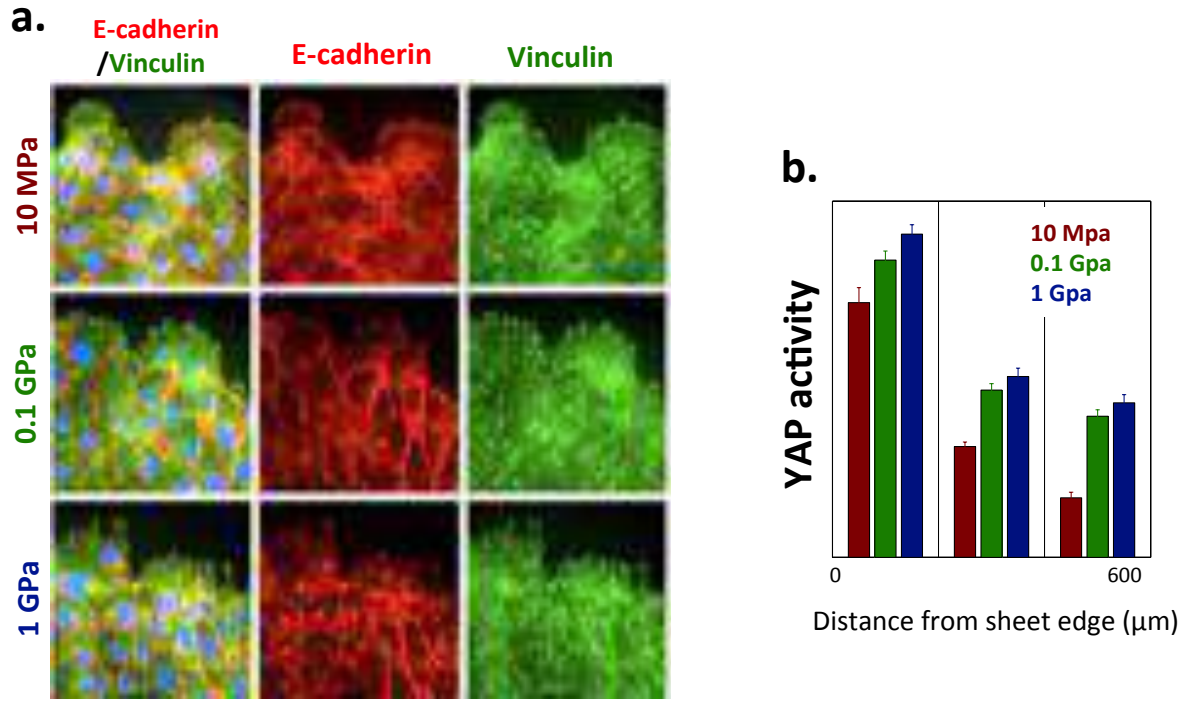

**Supplemental figure 7**

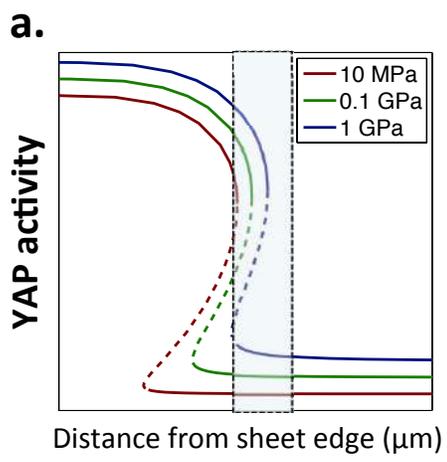 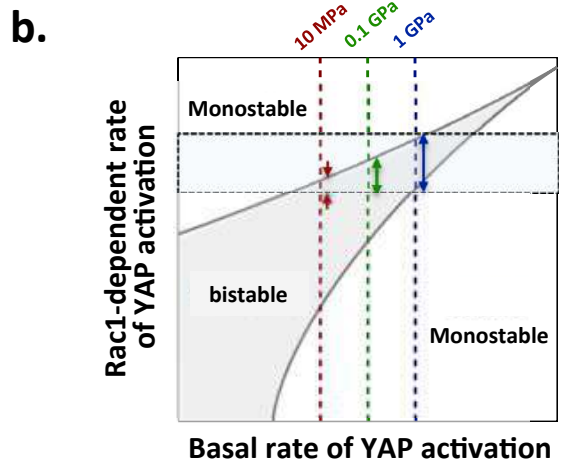

**Supplemental figure 8**

**Figure 1e**

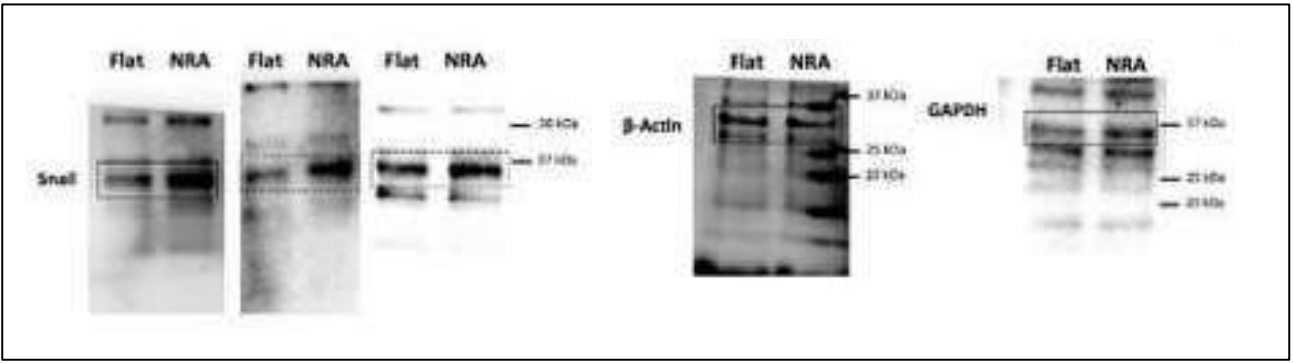

**Figure 1g**

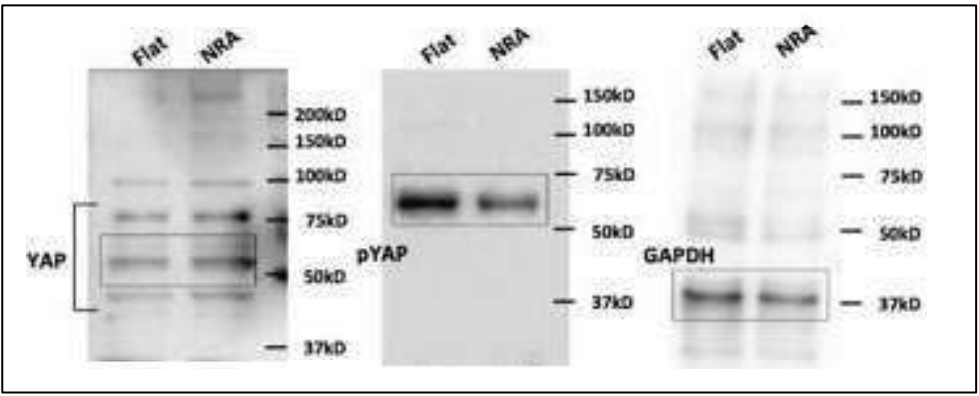

**Figure 2b**

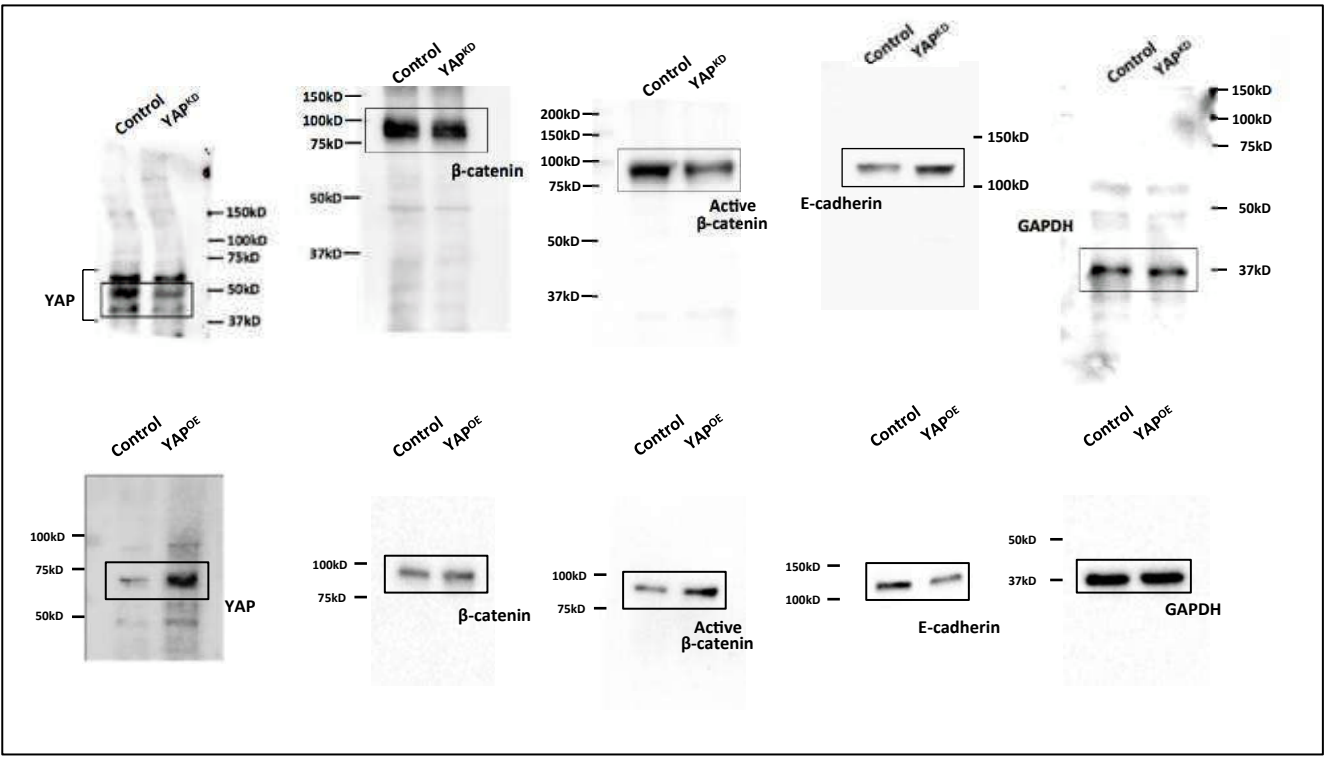

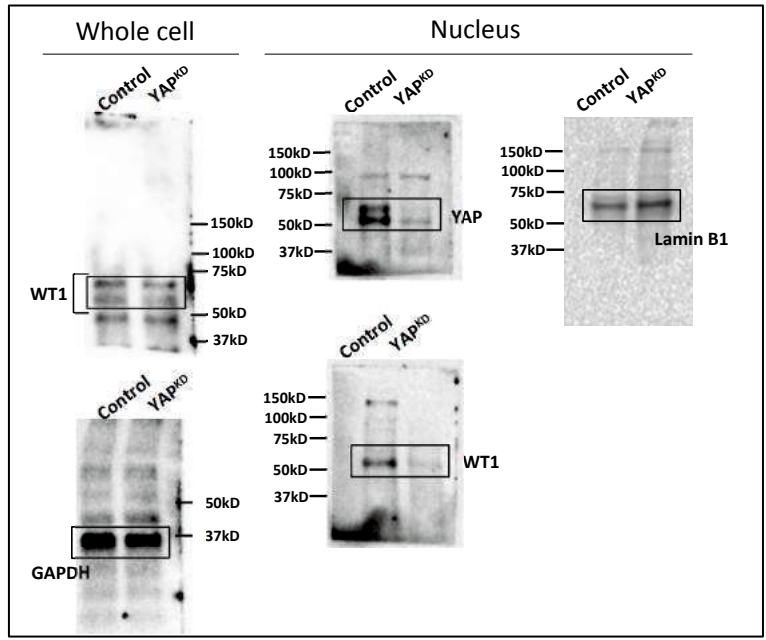
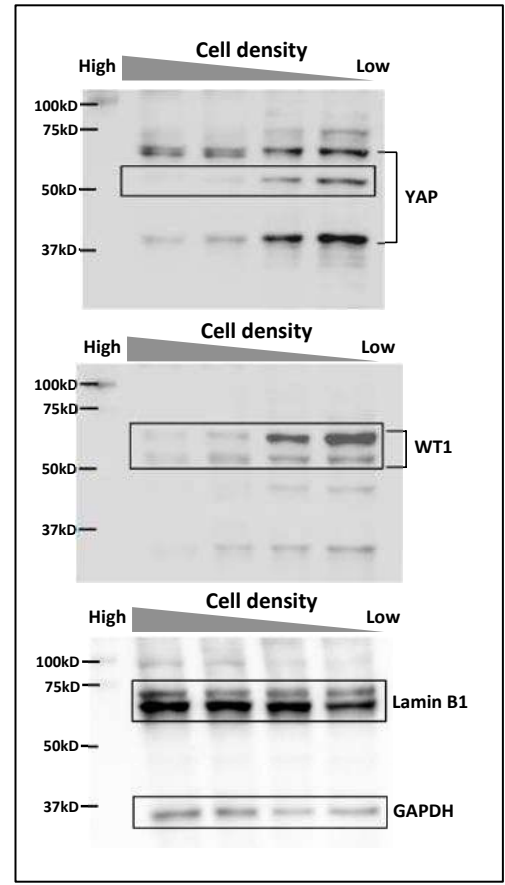
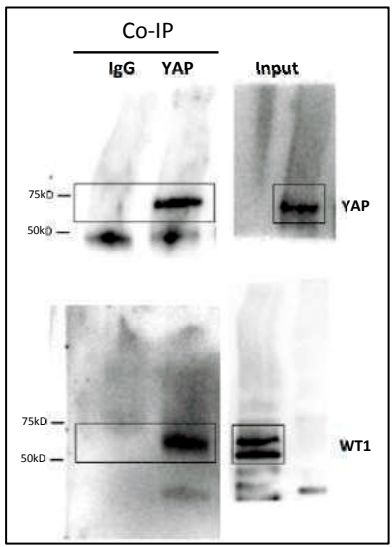
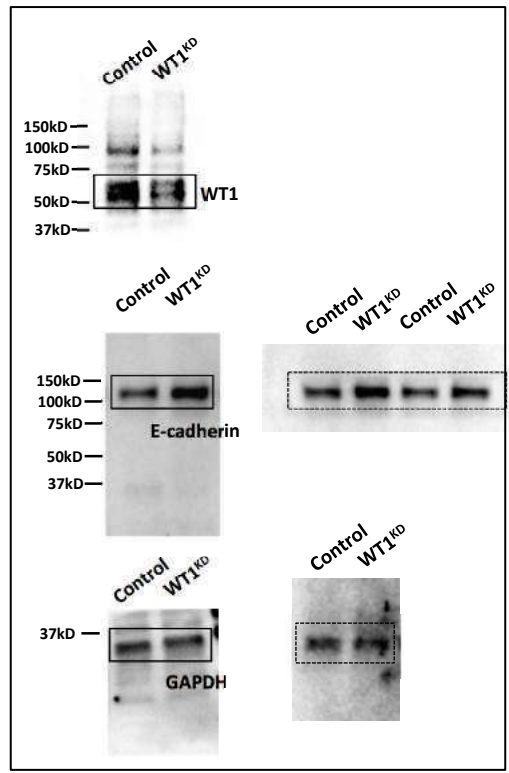
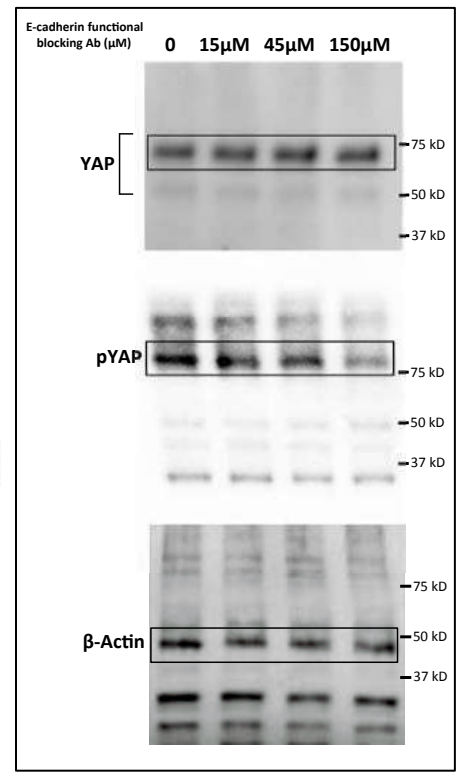

## Figure 4a

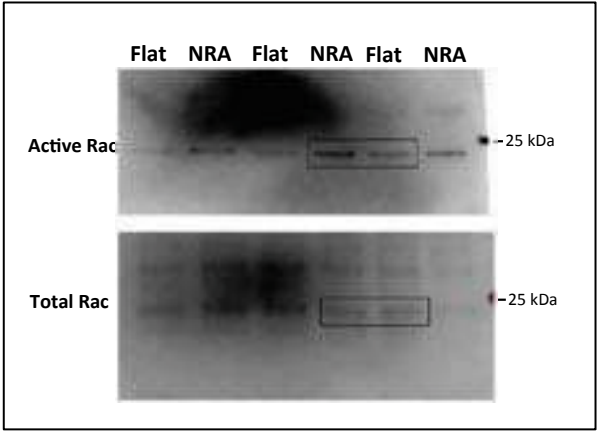

## Figure 4b

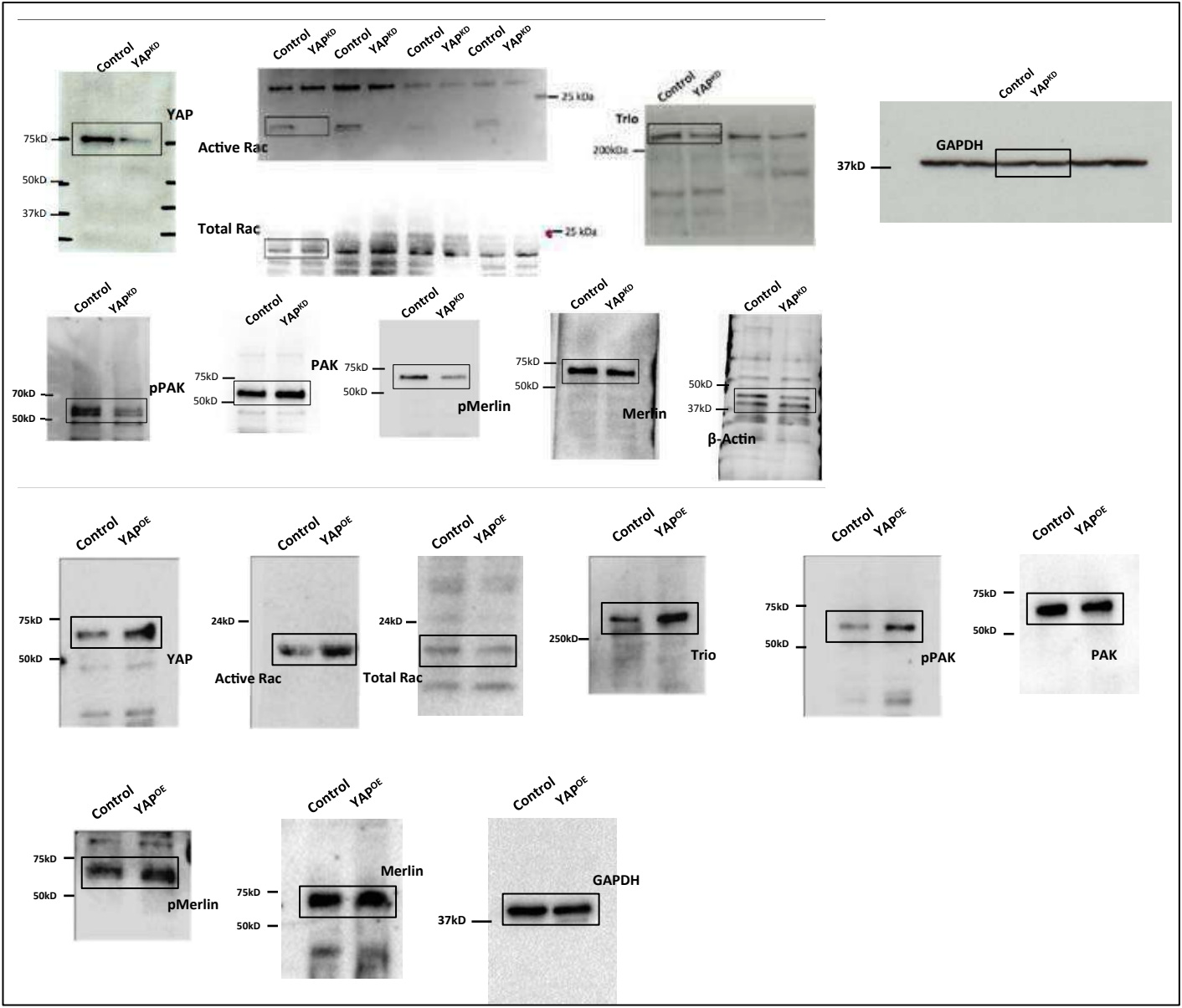

Figure 4e

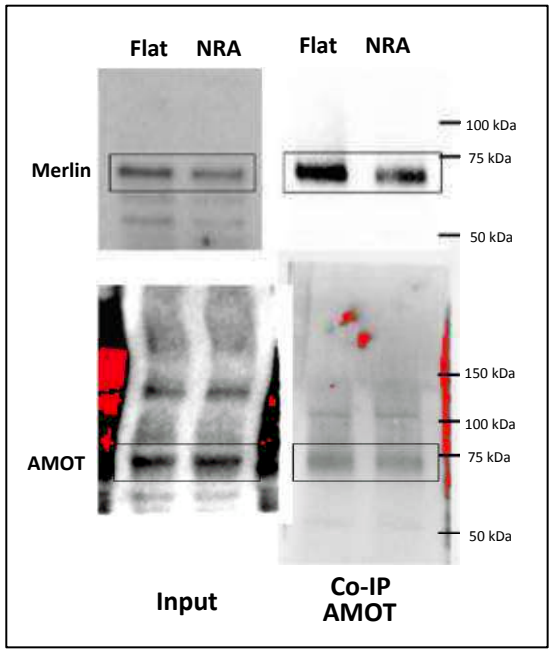

Figure 4f

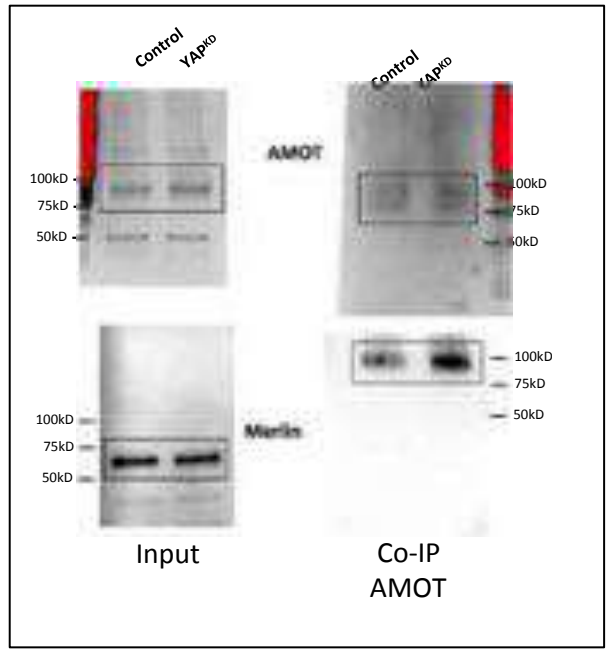

Figure 4g

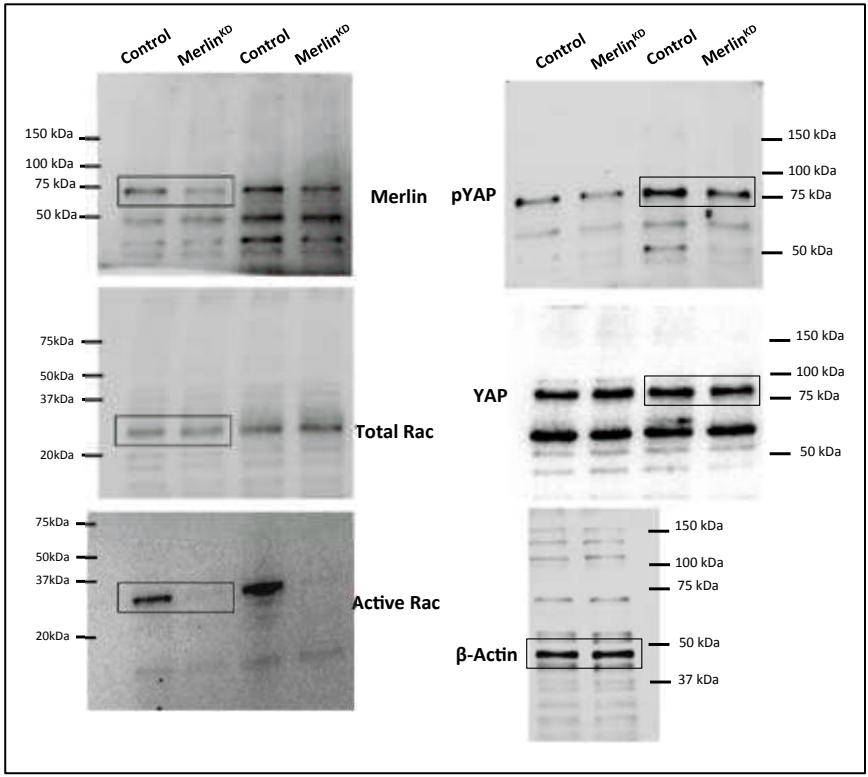

Supplemental figure 6

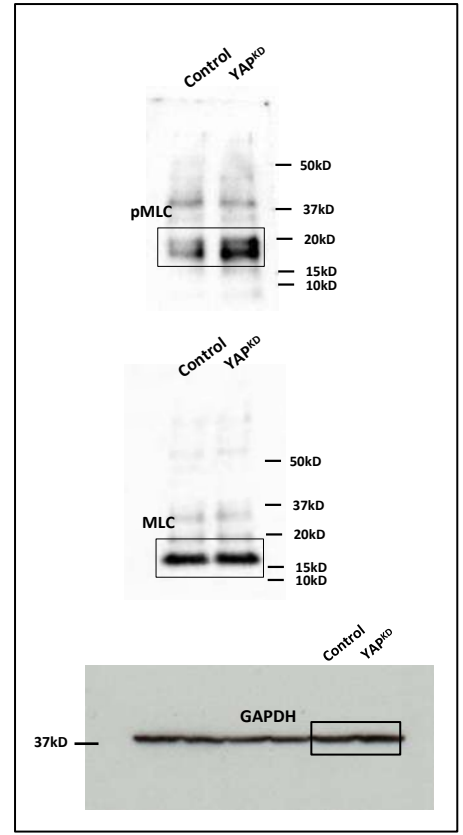

**Figure 6a**

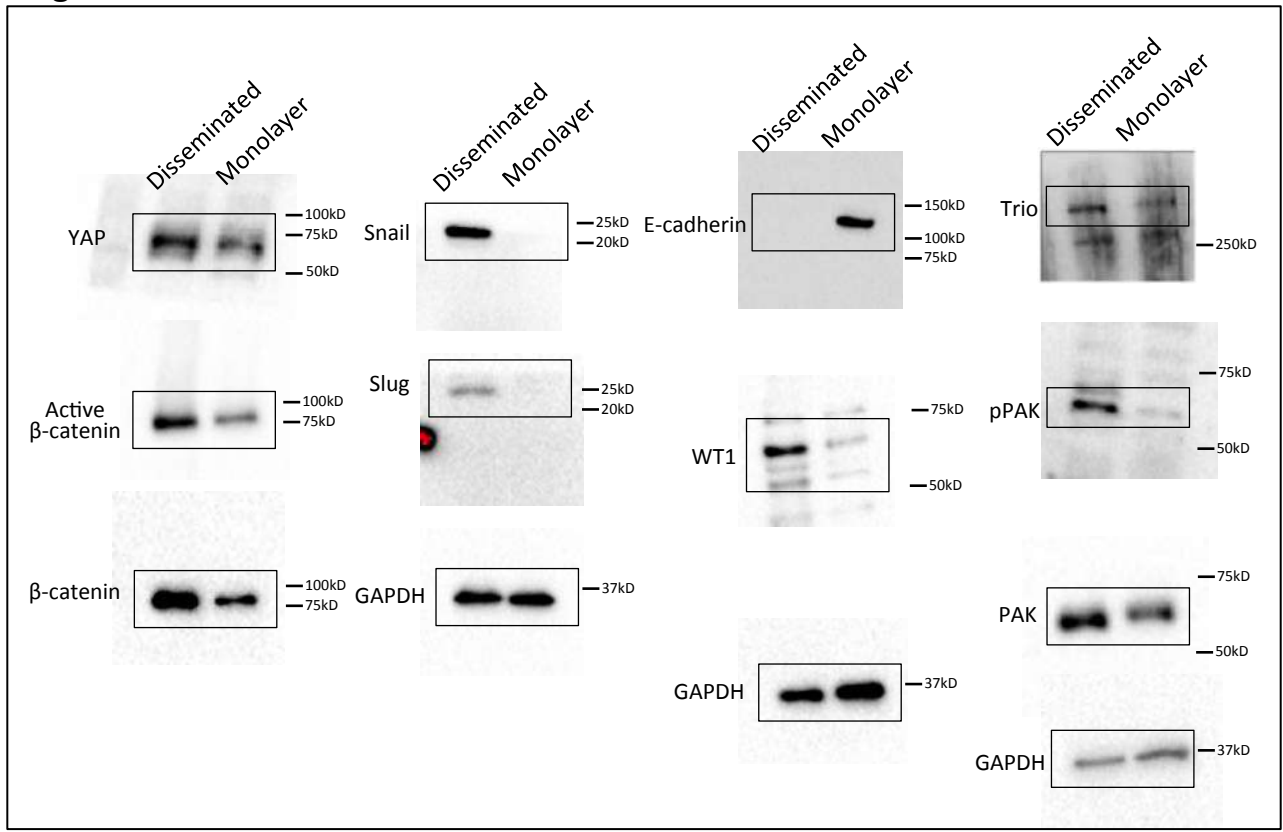